# Flow Instability Transferability Characteristics within a Reversible Pump Turbine (RPT) under Large Guide Vane Opening (GVO).

Maxime Binama[a], Kan Kan[b,*], Hui-Xiang Chen[c], Yuan Zheng[b,*], Daqing Zhou[b], Wen-Tao Su[d], Alexis Muhirwa[e], James Ntayomba[a]

[a]*College of Water Conservancy and Hydropower Engineering, Hohai University, Nanjing, 210098, PR China*
[b]*College of Energy and Electrical Engineering, Hohai University, Nanjing, 211100, PR China*
[c]*College of Agricultural Engineering, Hohai University, Nanjing, 210098, PR China*
[d]*College of Petroleum Engineering, Liaoning Shihua University, Fushun, 113001, PR China*
[e]*School of Energy Science and Engineering, Harbin Institute of Technology, Harbin, 150001, PR China*

***Abstract***: Reversible pump turbines are praised for their operational flexibility leading to their recent wide adoption within pumped storage hydropower plants. However, frequently imposed off-design operating conditions in these plants give rise to large flow instability within RPT flow zones, where the vaneless space (VS) between the runner and guide vanes is claimed to be the base. Recent studies have pointed out the possibility of these instabilities stretching to other flow zones causing more losses and subsequent machine operational performance degradation. This study therefore intends to investigate the VS flow instability, its propagation characteristics, and the effect of machine influx and runner blade number on the same. CFD-backed simulations are conducted on ten flow conditions spanning from turbine zone through runaway vicinities to turbine brake (OC1 to OC15), using three runner models with different blades (7BL, 8BL, and 9BL). While VS pressure pulsation amplitudes increased with runner blades number decrease, the continuously decreasing flow led to gradual VS pressure pulsation level drop within the Turbine zone before increasing to Runaway and dropping back to deep turbine brake zone. The effect of the same parameters on the transmission mode to VS upstream flow zones is more remarkable than the downstream flow zones.

1. Introduction

Due to caused substantial harm on the environment and so many other associated economic shortfalls, the utilization of fossil fuel as the main energy source has seen a tremendous fall in the last decades, while on the other hand, new renewable energy (NRE) sources have been increasingly adopted, leading to a recent high penetration rate of NREs within the electric power grid [1-3]. As much this practice provides an environmentally friendly energy generation means at the same time ensuring sustainability, the randomness associated with NREs generation still causes energy supply stability issues [4-6]. In this regard, owing to its high operational flexibility and the ability to store huge amount of energy, pumped-storage hydropower (PSH) scheme disposes of an undisputable potential to deal with NRE-caused grid instability[7-11]. PSH schemes are now widely used in different countries such China [12, 13], Brazil [14, 15], India [16, 17], Japan [18-20], United States [21, 22], and Europe (in general) [23-25]. Due to NRE production intermittency, the power grid may experience a mismatch between power supply and demand which consequently requires the PSH to provide the make-up power, thus ensuring safety and stability of power supply systems [26, 27]. However, due to these PSH operating circumstances, the utilized reversible pump turbines (RPTs) are forced into instant stops and start-ups, frequent switches between pumping and turbining modes, while being operated under off-design conditions for most of times [28, 29]. This, as also mentioned by different investigators, gives rise to huge flow instability within the machine water flow passages, which in turn comes with different associated detrimental phenomena such as pressure pulsations and subsequent structural vibrations, machine components breakage under serious cases, noise, efficiency degradation and difficult machine synchronization with the grid [30-32]. Note that off-design operating conditions-related flow instability has been recorded in almost all hydraulic machinery, including formal hydraulic turbines and pumps [33, 34]. While substantial instabilities have been spotted within the draft tube for hydraulic turbines, especially the Francis type [35-38], the most famous flow instability for RPTs that is constantly reported is situated within the vaneless space between the guide vane and the runner [39-41]. Indeed, as reported by Li et al. [42], a vaneless space water ring is formed under transient conditions, blocking the way to a continuous water flow, which as mentioned by many other researchers, Hasmatuchi et al. [43] and Yin et al. [44] among others, serves the trigger to an abrupt head increase, extreme high pressure pulsations, and corresponding formation of the s-shaped characteristic curve. This s-shape characteristics is however warned to not only take place when the vaneless space flow unsteadiness occurs, it's rather clarified that its emergence is specifically marked by the evolution of a fully developed rotating stall with a well-defined frequency [45]. Note that the rotating stall has been defined as the flow unsteadiness that leads to the emergence of rotating velocity pulsations of sub-synchronous nature [46]. Indeed as observed by different investigators [47-50], the developed rotating stall successively blocks a number of runner channels causing subsequent flow reversal, where it generally rotates with sub-synchronous speeds (50 to 70% of the rotating frequency). Xia et al. [51], investigating the RPT flow dynamics under off-design operating conditions, observed the occurrence of four circumferentially distributed stalls that blocked some of the flow channels, which in a long run led to the emergence of high pressure pulsations at same zones. This somehow agrees with the findings from works by Guggenberger et al. [52] and Zhang et al. [53], where it was confirmed that strong flow vortices accompanied by huge back flows emerged within the runner and guide vanes when the machine flow conditions reduced to turbine brake zone. This was accompanied by an abrupt increase in pressure pulsations, where low frequency components ($0.65f_n$) were spotted and have been attributed to the occurrence of cascade flow unsteadiness. Note also that, investigating the vaneless space pressure pulsation characteristics for RPTs under unstable flow conditions, the blade passing frequency component (BPF) and its harmonics have been observed, and attributed to the effect of rotor-stator interactions (RSI)[28, 54, 55]. The RSI has been explained by different investigators as the result of a combined effect of the pressure field that rotates with runner blades and the formed wakes behind every guide vanes [56, 57]. This phenomenon has been found to be the source of strong pressure pulsations that, under specific operating conditions, can even propagate to other flow zones [58]. Among the most recently published details on RPT flow instability, Ma et al. [59] recorded high vaneless space pressure pulsations with the above mentioned BPF components and its harmonics as well as the low frequency components (LFCs). The later were found to be linked to flow instabilities that took place when the machine operating conditions were within the s-shape operating zone. In this study, the blade lean angle was found to considerably influence the machine flow structure formation mechanism as well as the associated pressure pulsation distribution mode. On the other hand, an RPT draft tube vortex rope was investigated by Lai et al. [60], where through the analysis of draft tube tangential and axial flow velocity profiles, deep part-load conditions were found to be accompanied by the emergence of draft tube strong swirling flows that rotated in the same direction as the runner, while the contra-rotating swirl type was observed for high load conditions. The draft tube rotating vortex rope frequency was estimated to be $0.28f_n$ ($f_n$: runner rotational frequency). This is somehow similar to the case of Francis turbine's vortex rope where for instance, as found by Zhao et al. [61] and Duparchy et al. [62], the cavitating draft tube vortex rope procession frequency was within the range of $0.25f_n$ to $0.35f_n$. Zhang et al.[63], studying the RPT flow dynamics for a wide range of flows, observed the occurrence of the draft tube vortex rope under turbine brake conditions, where draft tube pressure pulsations were confirmed to be the result of a combined effect of the vaneless space rotating stall and the draft tube vortex rope. More details about RPT transient manoeuvers that instigate flow unsteadiness emergence such as load rejection, machine start-up, pump-trip, etc.… were recently published adding to a lot more findings from the past decades. Among others, Mao et al. [64] investigated the RPT flow instability and induced noise for a machine through transient load rejection process by continuous and intermittent guide vanes closure. The acoustic pressure radiation characteristics were consistent with eventual development of the machine flow dynamics, which makes this technics adequate for flow instability prediction. This kind of instability had also been explored by Li et al. [65], considering RPT's continuous guide vanes closure under pump mode. Serious pressure and velocity pulsations took place within the inter-guide and stay vanes flow zones, especially in the last phases of the GVO closure process. This was found to mainly take source from the emerged flow vortices that occupied more channels as the guide opening angle continuously reduced. On the other hand, a description of the flow structure evolution mechanism and associated pressure pulsation changes for a reversible pump turbine through the pump-trip transient process has been presented by Yang et al. [66]. It was found that, depending on the instant changes in machine flow conditions, local flow structures constantly changed, leading to the formation of different secondary flows under specific conditions. These are for instance the rotating stall within the inter-blade and the inter-vane flow channels, impinging flow jets on blade pressure surfaces, and vortex rope within the draft tube. The appearance, evolution, and vanishment of these structures at different flow zones led to the emergence of high pressure pulsations and uneven distribution of runner force pulsations. Note that, in a corresponding way, in a study conducted by Li et al. [39] back in 2016, radial forces that are perpendicular to the runner shaft have been found to emerge only under transient conditions such as the vicinities of runaway and turbine brake zones. These were also found to be linked to the emergence of unevenly distributed vortices and backflows within the runner inter-blade channels and vaneless space under these conditions.

Taking reference from the above-presented details and many others from different journals and other sources, the already carried out investigations have confirmed the vaneless space between the runner and guide vanes to be the base for RPT flow instability. In addition, it has also been revealed that, depending on the machine operating conditions, these instabilities can stretch to other flow zones in both the up and downstream sides. For instance, in a study conducted by Widmer et al.[50], circumferentially distributed stationary vortices emerged within the runner and vaneless space under low GVOs, while a rotating stall was noticed within the same zones, but now with a possibility to extend to upstream flow zones for large GVOs. In this respect, targeting a thorough understanding of RPT flow structures formation mechanism and associated pressure field characteristics, especially under transient operating conditions, the present study seeks to deeply investigate the flow instability transferability characteristics between the vaneless space zone and both the upstream and downstream flow zones for an RPT operating under large guide vane opening (GVO: 34mm), and devise the effect of runner blade number on the same. In this study, 3D numerical simulation of the RPT flow has been conducted for ten flow conditions (C1 to C10) expanding the whole range of turbine, runaway vicinities, and turbine brake zones. Moreover, three runner models with 7, 8, and 9 blades have been utilized to investigate the effect of runner design on flow parameters transferability among different components of the concerned computational domain. Numerical and experimental results in terms of Discharge-Speed ($Q_{11}$-$n_{11}$) characteristic curve have been first compared for validation of the utilized numerical simulation scheme. Next, a deep analysis of the RPT flow and pressure fields characteristics under different operating conditions has been carried out, where a number of concluding remarks were finally drawn.

## 2. Computational domain and method
### 2.1. Computational domain

The numerically investigated geometric model, is a low specific speed Francis type reversible pump turbine, composed of five main components, namely the volute casing, stay vanes ring, guide vanes ring, runner, and draft tube. The number of stay vanes, guide vanes, and runner blades are 20, 20, and 9 respectively. Dimensions of the here-investigated reduced-scale RPT model are presented in Table 1.

Table1. Geometric parameters of the investigated RPT model.

| Parameter | Symbol | Value |
| --- | --- | --- |
| Runner Specific Speed (min$^{-1}$) | $n_q$ | 36.8 |
| Runner Inlet Dia. (mm) | $D_2$ | 560 |
| Runner Outlet Dia. (mm) | $D_1$ | 270 |
| Runner Blade Number (-) | $Z_R$ | 9 |
| Guide Vane Distribution Dia. (mm) | $D_0$ | 662 |
| Guide Vane Height (mm) | $B_0$ | 37.8 |
| Guide Vane number (-) | $Z_0$ | 20 |
| Stay Vane Number (-) | $Z_S$ | 20 |
| Stay Vane In Dia. (mm) | $D_{SI}$ | 966 |
| Stay Vane Out Dia. (mm) | $D_{SO}$ | 763 |

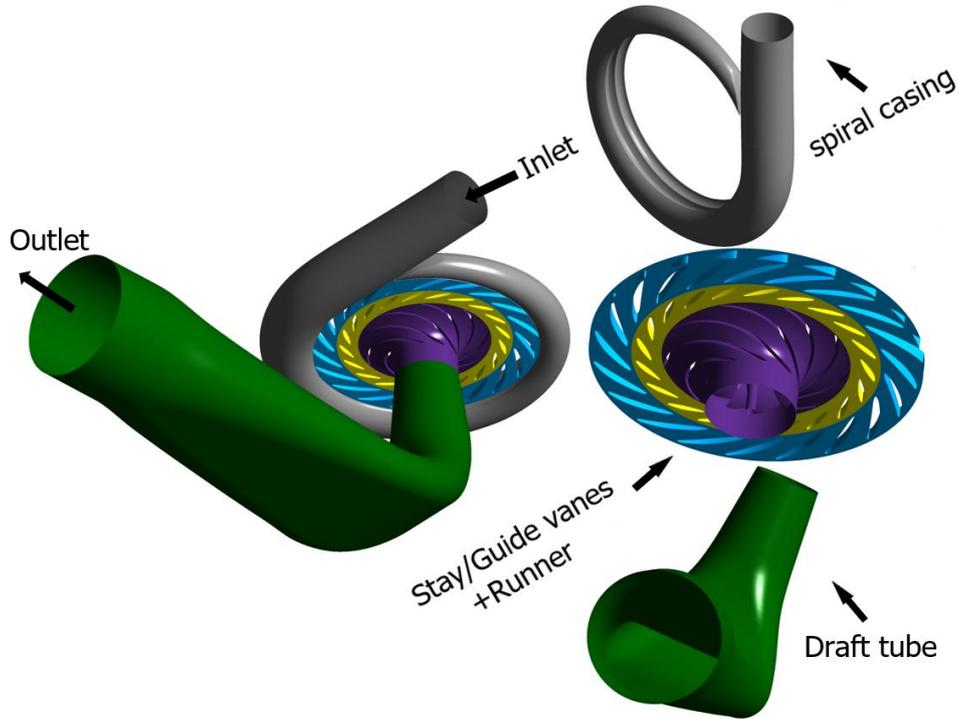

Fig. 1 RPT computational domain and components

The whole computational domain was built using the geometrical design software Unigraphix NX 8.0. The RPT geometric model and components are presented in Fig. 1. Experimental tests were carried out on the here-presented model, where hydraulic machinery testing standards as set through IEC [67] have been neatly followed. Note that the turbine operating mode of RPT involves the water entering through the volute casing's inlet, then successively passes through stay and guide vanes inter-spaces respectively, before whirling through the runner inter-blade flow channels and exit the testing section through the draft tube's outlet zone. The same water flow trajectory is adopted for numerical simulations. Though the experimental tests on RPT performance characteristics have been conducted on a wide range of flow conditions and guide vane openings (1mm to 39mm GVO), the present study only focuses on one opening, namely GVO: 34mm. Both Fig. 2 and Fig. 3 show the RPT experimental testing section and the schematic view of the utilized testing system, while Fig. 4 shows the experimentally obtained RPT characteristic curve $Q_{11}$-$n_{11}$. Note that, as shown in Fig. 4, the s-shaped trend of the RPT characteristic curves for different GVOs is obvious mostly at higher values of rotational speed for all tested GVOs.

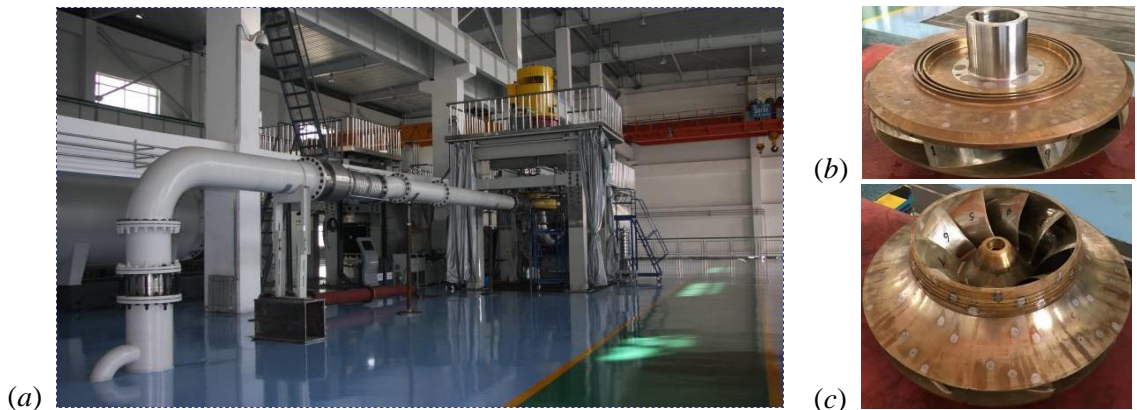

Fig. 2 RPT test rig (a) testing section (b and c) tested runner up and downside.

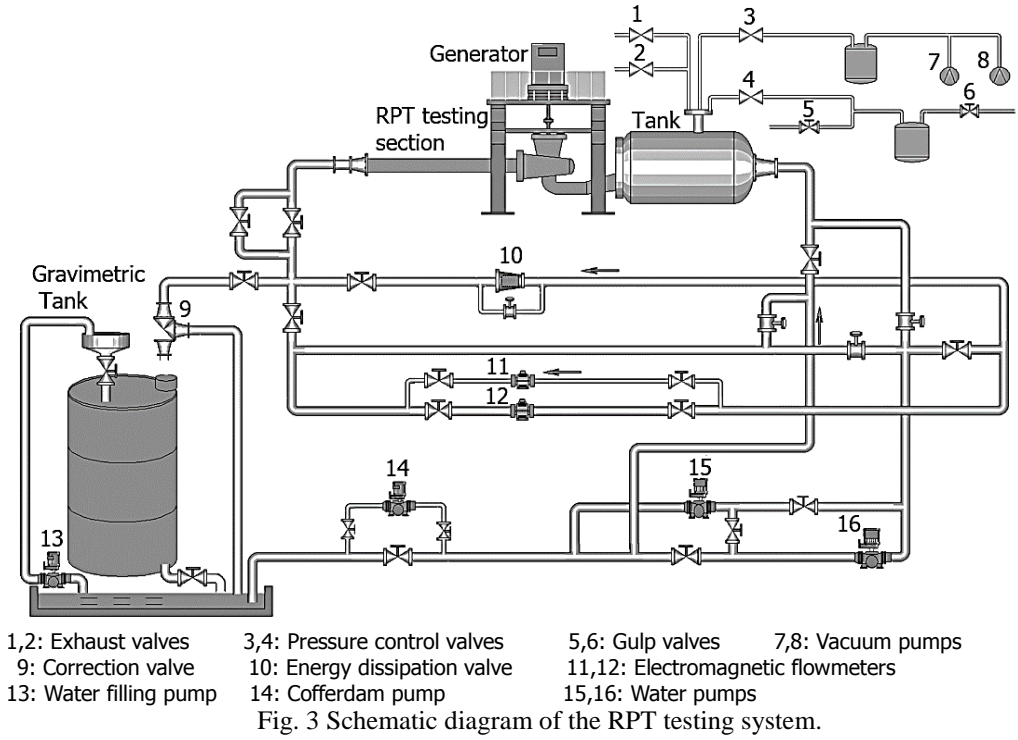
1,2: Exhaust valves  3,4: Pressure control valves  5,6: Gulp valves  7,8: Vacuum pumps
9: Correction valve  10: Energy dissipation valve  11,12: Electromagnetic flowmeters
13: Water filling pump  14: Cofferdam pump  15,16: Water pumps

Fig. 3 Schematic diagram of the RPT testing system.

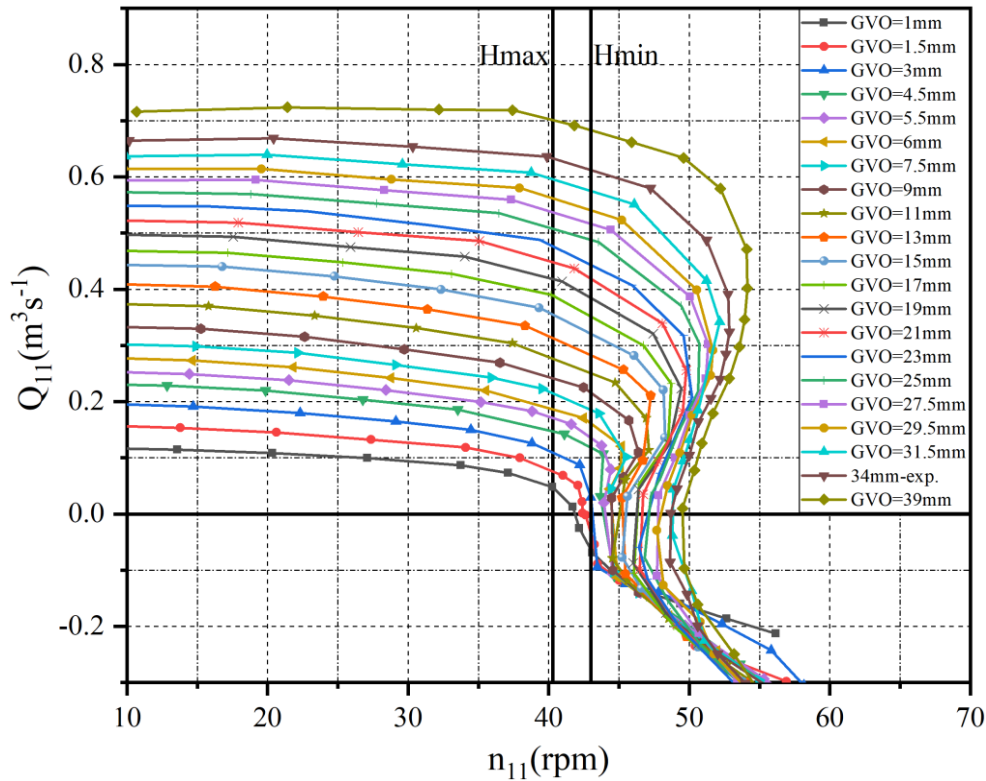

Fig. 4 Experimental $Q_{11}$-$n_{11}$ graph.

## 2.2. Computational method
### 2.2.1. Grid generation

In line with the present study's main objectives, the computational grid for the RPT flow domain encompassing the full RPT flow passage had to be generated, as the preparatory phase for followed numerical simulations. Therefore, the computational grid for the above-stated five RPT components has been generated using two Ansys mesh generation codes, namely Ansys ICEM and Ansys Turbogrid. The structural hexahedral grid type has been generated for the majority of components, while the unstructured tetrahedral grid type has been used at the volute casing's tongue, owing to its sharp geometry. In order to quite accurately capture the flow dynamics at critical flow zones such as the blade/vane walls, much finer grid has been particularly generated at those areas, where y+ value has been globally below 30. Note that this value is generally used to express the distance of the grid's first node to the wall. Moreover, in order to avoid the dependence of numerical simulation results on the utilized grid number, eight different grid numbers ranging from 2.4 to 12.3 million elements have been generated and separately tested under similar boundary conditions. Taking the RPT head (H) as the testing parameter as shown in Fig. 5, the H value first quickly decreased as the grid number increased before almost stabilizing for the last three values. Therefore, in line with the available computational resources, the 9.3 million grid number has been considered for farther simulations. Table 2 shows the details of the selected grid, while Fig. 6 gives an external view of the generated grid for certain components.

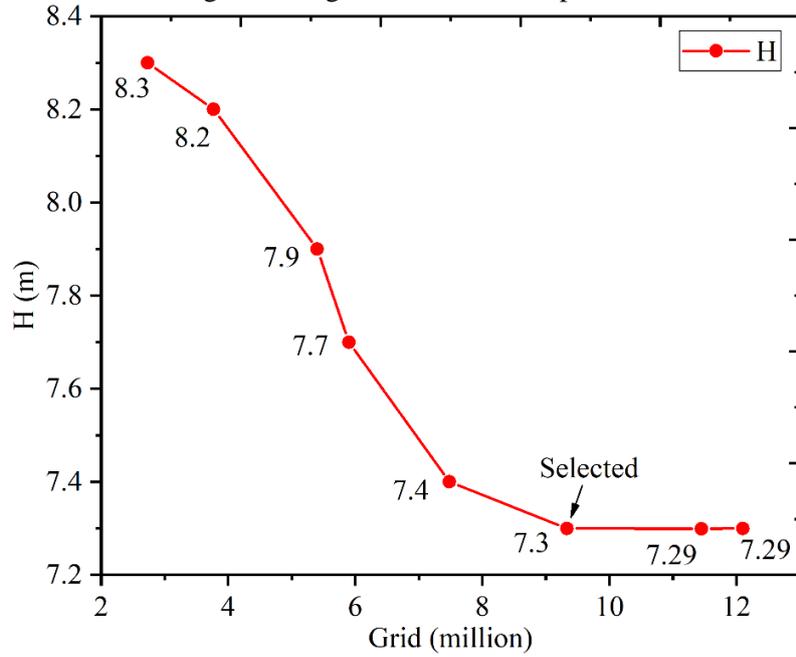

Fig. 5 RPT grid independence test graph.

Table 2 Selected grid details for different RPT components

| Components | Grid type | Grid number (million) | Quality |
|---|---|---|---|
| Spiral casing | Hexahedral+tetra | 0.8 | 0.3 |
| Stay vanes | Hexahedral | 1.2 | 0.6 |
| Guide vanes | Hexahedral | 1.1 | 0.6 |
| Runner | Hexahedral | 4.6 | 0.4 |
| Draft tube | Hexahedral | 1.6 | 0.7 |
| Total | - | 9.3 | - |

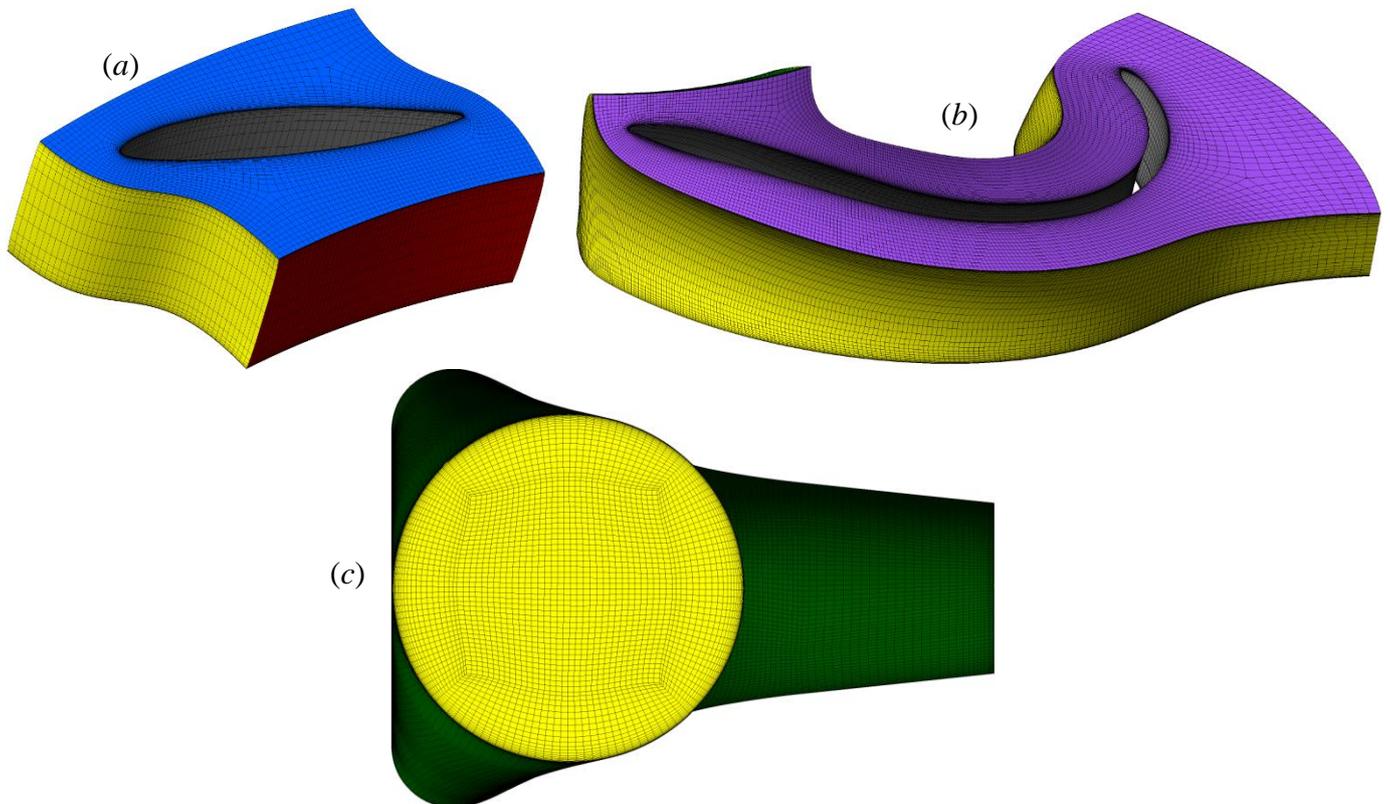

Fig. 6 A glimpse on structural hexahedral grid in (a) guide vanes, (b) runner blades, and (c) draft tube.

*2.2.3. Numerical scheme*

Numerical simulations are conducted on a RPT operating under different flow conditions. The CFD commercial code Ansys CFX has been used for the accomplishment of that task. Moreover, the Shear Stress Transform (SST) model is adopted for the closure of the flow-governing RANS equations. Note that these equations are a set of mass and momentum conservation equations that guide the dynamics of different fluids (see Equations 1 and 2 in Cartesian form).

The here-used SST turbulence model is known to combine the advantages of its two predecessors, namely $k$-$\varepsilon$ and $k$-$\omega$ models, thus providing more potential to quite accurately capture the details of RPT flow dynamics both in wall vicinities and within the flow bulk. Its mathematical expressions are given in Equations 3 and 4.

$$\frac{\partial C'_i}{\partial x_i} = 0 \tag{1}$$

$$\frac{\partial \bar{C}_i}{\partial t} + \bar{C}_j \frac{\partial \bar{C}_i}{\partial x_j} = \frac{1}{\rho}\left(-\frac{\partial \bar{p}}{\partial x_i} + \mu \frac{\partial^2 \bar{C}_i}{\partial x_j^2}\right) - \frac{\partial \overline{C'_i C'_j}}{\partial x_j} + f_i \tag{2}$$

$$\frac{\partial(\rho k)}{\partial t} + \frac{\partial(\rho u_j k)}{\partial x_j} = P - \beta^* \rho \omega k + \frac{\partial}{\partial x_j}\left[(\mu + \sigma_k \mu_t)\frac{\partial k}{\partial x_j}\right] \tag{3}$$

$$\frac{\partial(\rho \omega)}{\partial t} + \frac{\partial(\rho u_j \omega)}{\partial x_j} = \frac{\gamma}{\nu_t} P - \beta \rho \omega^2 + \frac{\partial}{\partial x_j}\left[(\mu + \sigma_\omega \mu_t)\frac{\partial \omega}{\partial x_j}\right] + 2(1-F_1)\frac{\rho \sigma_{\omega 2}}{\omega}\frac{\partial k}{\partial x_j}\frac{\partial \omega}{\partial x_j} \tag{4}$$

Throughout the RPT flow simulation process, steady state numerical simulations were first carried out, the results of which served as initial state of the followed transient simulations. While 21 guide vane openings (1 to 39mm GVO) were experimentally tested, this study limits itself on only one GVO value of 34mm. In addition, among all the operating conditions associated with the 34mm GVO, <u>Ten operating conditions</u> (OC1 to OC15) spanning all the way from turbine zone, through runaway vicinities, to turbine brake were selected for farther investigation (see Table 3). It's also worth mentioning that the original RPT model's runner had 9 blades as shown in Table 1. However, in order to investigate the effect of runner geometry on flow instability transmissibility among RPT components, which constitutes one of this study's objectives, the blade number has been decreased from 9 through 8 to 7, thus resulting in three runner models in total, namely BN7, BN8, and BN9. Throughout the whole simulation process, values for crucial boundary conditions such as the inlet flow rate $Q_{in}$, outlet static pressure $P_{out}$, and runner rotational speed $n$, have been picked from correspondent experimental data. The whole RPT computational domain was divided into stationery and rotating frames, between which, different interfaces were utilized depending on the case. For steady state numerical simulations, the "frozen rotor-stator" interface type was used between the runner and downstream flow zone, while the "stage mixing" interface type was used between the runner and upstream flow zones. For transient numerical simulations however, the "transient rotor-stator" type has been used on both runner ends. The non-slip wall condition was imposed on all walls. For all transient simulation sessions, 10 runner revolutions were considered for each, where the adopted time-step Δt was equivalent to 1 degree runner rotation, making ten revolutions worth 3600 time-steps. In addition, 5 internal loops have been selected for every time-step, while the residuals Root-Mean Square bellow $1\times10^{-05}$ was selected as the solution convergence criterion.

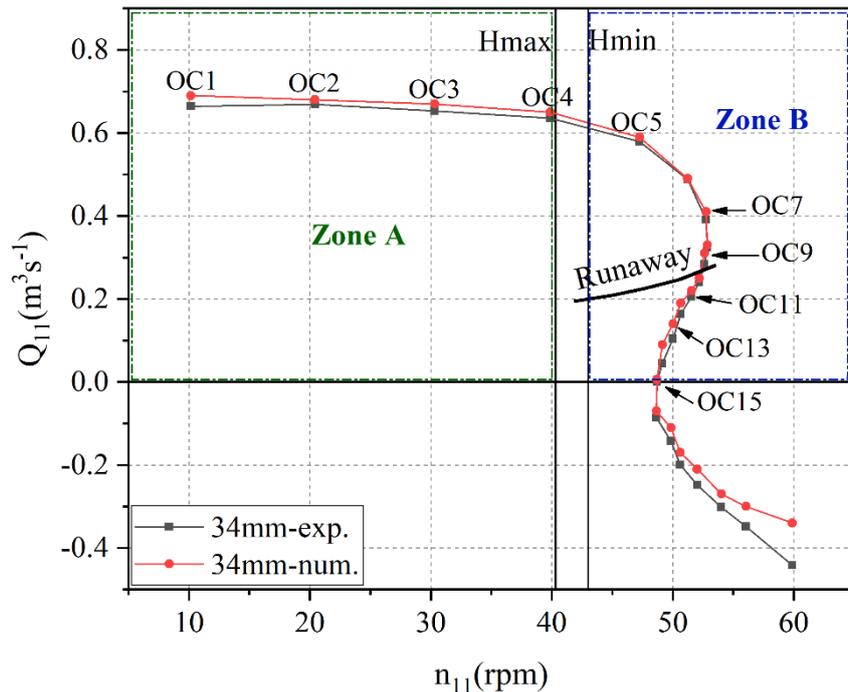

Fig. 7 Comparison of numerical and experimental Discharge-Speed ($Q_{11}$-$n_{11}$) curve

Table 3 Details about the ten investigated operating conditions

| No. | Operating conditions (OCs) | $Q_{11}$(m$^3$s$^{-1}$) | $n_{11}$ (rpm) | Operating zones |
|---|---|---|---|---|
| 1 | OC1 | 0.664269 | 10.166348 | |
| 2 | OC2 | 0.668483 | 20.429189 | |
| 3 | OC3 | 0.653346 | 30.326086 | Turbine |
| 4 | OC4 | 0.635795 | 39.877927 | |
| 5 | OC5 | 0.579705 | 47.254189 | |
| 6 | OC7 | 0.390639 | 52.759861 | |
| 7 | OC9 | 0.283909 | 52.614695 | Runaway vicinities |
| 8 | OC11 | 0.205483 | 51.54634 | |
| 9 | OC13 | 0.104235 | 50.014213 | Turbine brake |
| 10 | OC15 | 0.00129 | 48.690881 | |

In order to validate the above simulation scheme's trustworthiness, numerically found results in terms of $Q_{11}$-$n_{11}$ characteristic curves have been compared to experimental ones, where the error between them was less than 4% on a global scale, thus making the utilized numerical scheme and its results quite trustworthy (see Fig.7). Note that the mathematical expressions of unitary Rotational speed $n_{11}$, Discharge $Q_{11}$, and Torque $T_{11}$ are presented in Equations 5, 6, and 7 respectively. It's also important to mention that the utilized values for Head and Torque are the averages of numerically found variations for each parameter, where among the simulated ten runner revolutions, the extracted results for farther analysis were only for the last four revolutions.

$$n_{11} = \frac{nD_1}{\sqrt{H}} \quad (5)$$

$$Q_{11} = \frac{Q}{D_1^2 \sqrt{H}} \quad (6)$$

## 3. Results and discussion

### 3.1. Flow field characteristics and analysis

In order to investigate the flow dynamics and eventual changes of structures within the machine cascade flow channels for the ten investigated flow conditions, a mid-span plane passing through stay/guide vanes and runner flow zones was considered as shown in Fig. 8. Considering the span-wise distance from hub to shroud, the hub and shroud positions are considered to be at 0 and 1 respectively, while the mentioned mid-span plane's position is at 0.5. In the same figure, the positions of stay/guide vanes, the vaneless space, and runner blades are shown. Fig. 9 displays the eventual changes in flow velocity profile as well as the corresponding flow vorticity dynamics within the cascade, as the machine flow conditions gradually decreased from the turbine zone (OC1-OC7), through runaway vicinities (OC9, OC11), to turbine brake zone (OC13, OC15). Under OC1 conditions, the machine influx was the 2$^{nd}$ highest ($Q_{11}$=0.6643m$^3$/s), while the runner rotational speed is the lowest ($n_{11}$=10.2rpm). These two parameters increased to OC2 conditions, where the discharge increased by a bit (0.6%) while the value of the runner rotational speed almost more than doubled. As it can be seen from Table 3 and Fig. 7, the undergone gradual flow decrease between every two successive flow conditions within the first phase of the turbine zone (OC1 to OC4) were of small values ($\Delta Q \leq 0.005$). However, this trend changed to a steep one for the last phase of the turbine zone (OC4 to OC7), turning even steeper for the followed operating conditions within the runaway vicinities and turbine brake zones, with the last operating condition being very close to no-flow condition ($Q_{OC15}$=0.00129m$^3$/s), below which the reverse turbine zone starts. On the other hand, the runner rotational speed almost constantly increased within the turbine zone ($5.5 \leq \Delta n \leq 10$), while it decreased within runaway vicinities and turbine brake zone, marking the s-shaped trend of the machine flow-speed curve ($Q_{11}$-$n_{11}$ curve). This same zone is believed to be characterized by large flow separations and high pressure pulsations within the machine flow channels. The velocity and vorticity contours for all tested conditions as shown in Fig. 9 showcase great flow structure changes within the cascade as the machine flow discharge $Q$ gradually decreased.

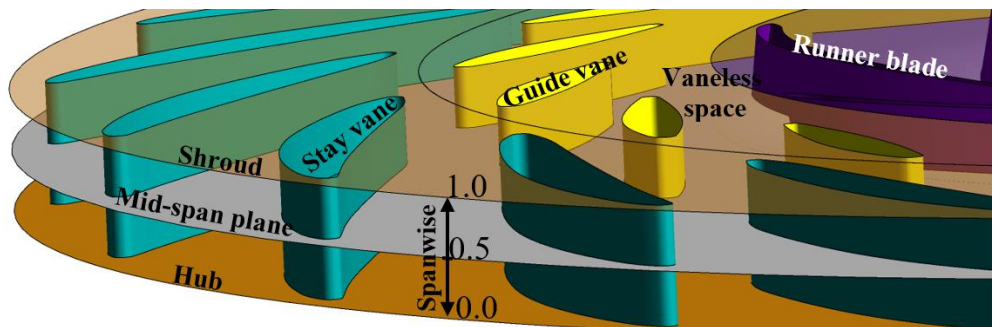

Fig. 8 Mid-span plane presentation

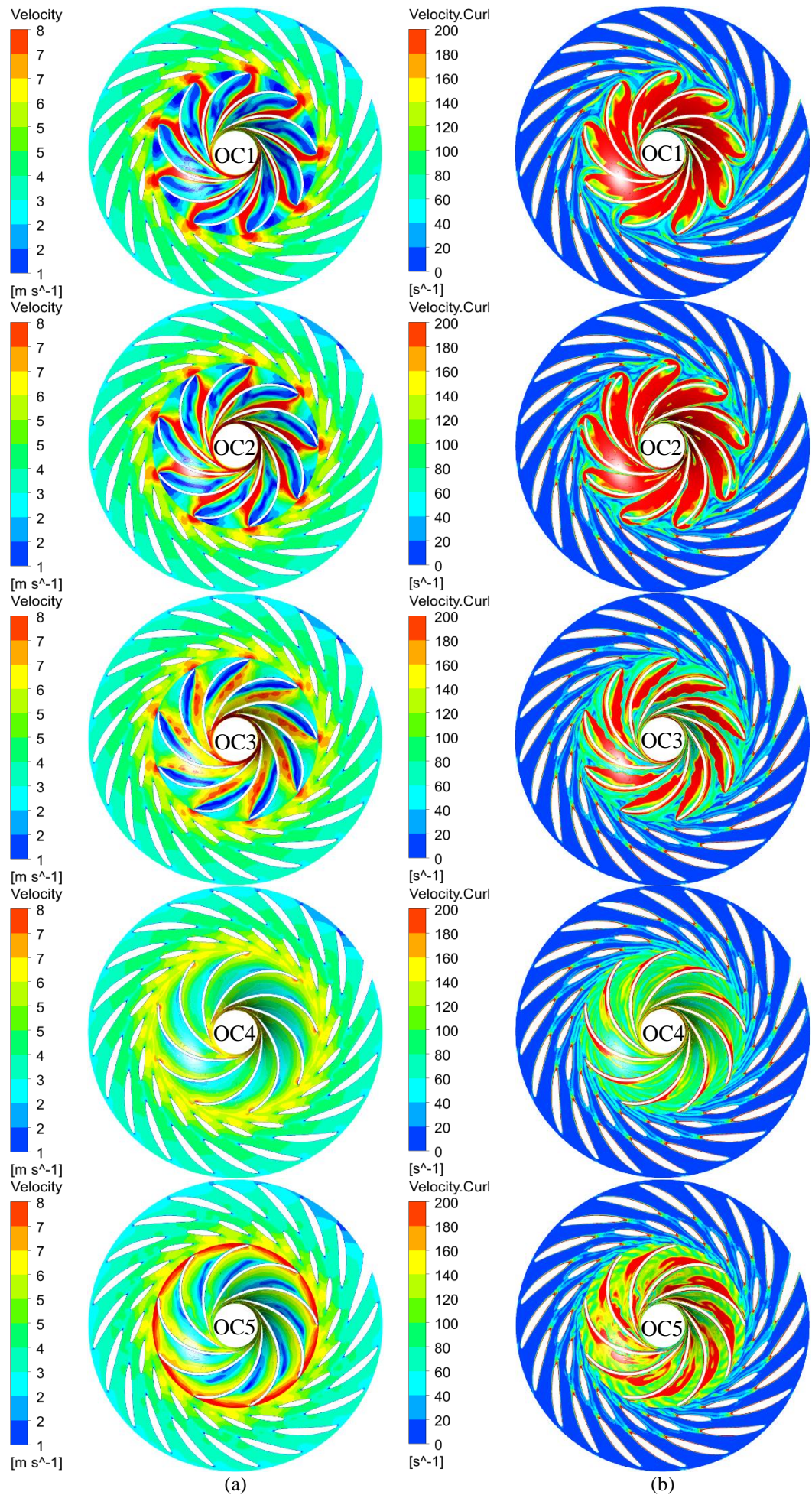

Fig. 9 Cascade flow field characteristics for different operating conditions (a) velocity contours (b) velocity curl contours

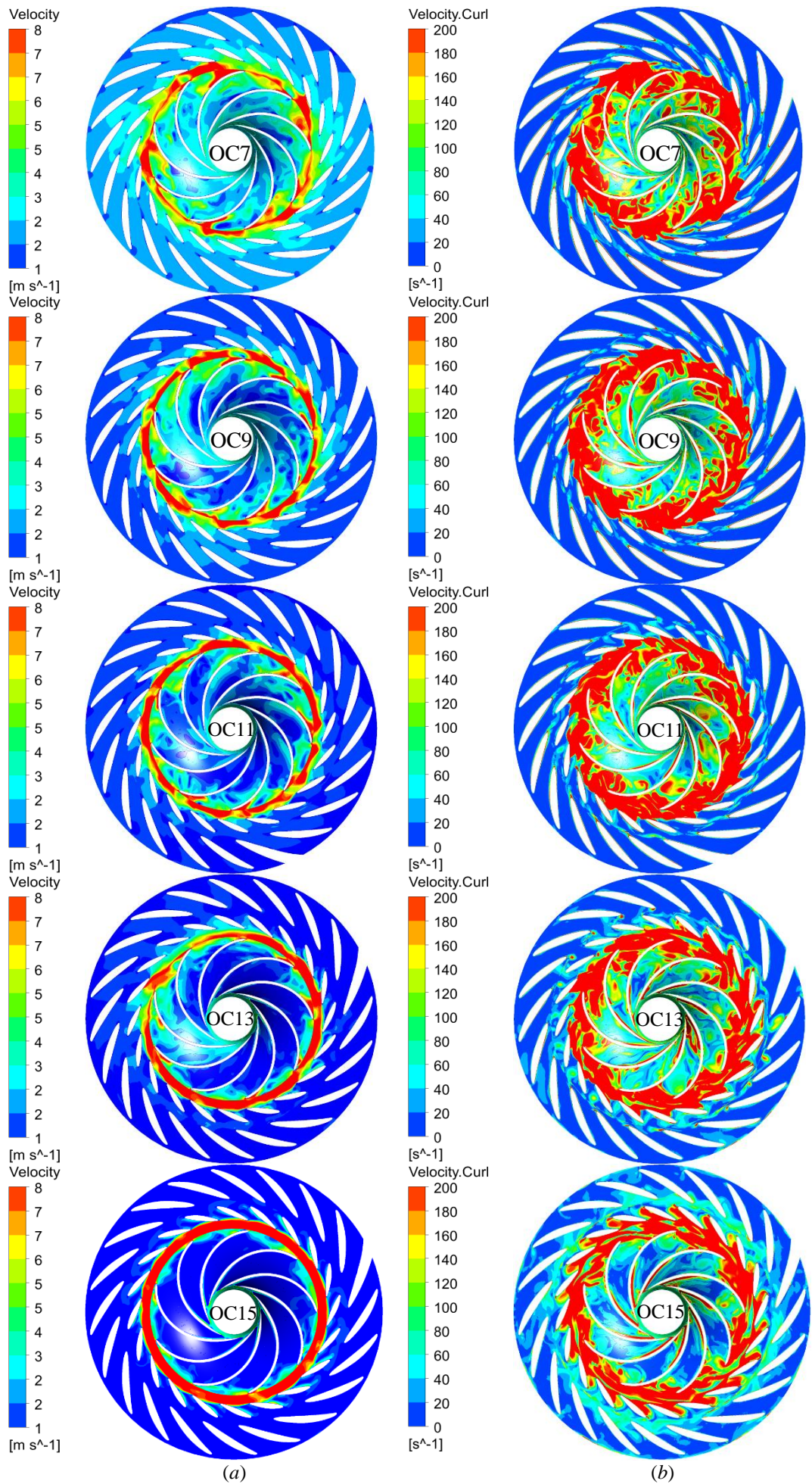

Fig. 9 (cont.)
Under OC1 conditions, the machine influx is still one of the highest, leading to a larger incidence angle at the runner blade leading edge (runner inlet zone). Under these conditions, due to higher flow speed within cascade flow channels, strong flow wakes have emerged at every stay and guide vanes trailing zone, where wakes behind guide vanes can even stretch to runner inter-blade flow channels. Due to the experienced water impinging on runner blade's pressure side (BPS) in the vicinities of the leading edge (BLE), the flow stagnation zone is formed at impinged surface, from where flow separation occurs leading to two components flowing in two opposite directions: the BPS-attached water flow component towards the runner deep inter-blade flow zones and the BPS-attached backflow component towards the vaneless space (VS) flow zone (see Fig.11). The later arrived at the VS zone, collides with both the VS circumferential flow and inter-guide vane flow, and form a flow obstruction between the runner BLE and the guide vane trailing edge (GVTE), leading to the jet flow formation at every runner BLE. The BLE jet flow is constituted by high speed flow streams that extend towards the runner inter-blade flow channels, eventually colliding on the BPS at the vicinities of the mid-distance blade streamwise length. At this zone, the BLE-born jet flow meets with the formerly mentioned flow component from the upstream BPS flow separation to form a high flow speed zone in the vicinities of the BPS. The subsequent pressure gradient between flow zones in the vicinities of the pressure side (BPS) and suction side (BSS) of two successive blades leads to the emergence of large BSS-attached vortices within the runner inter-blade flow channels.

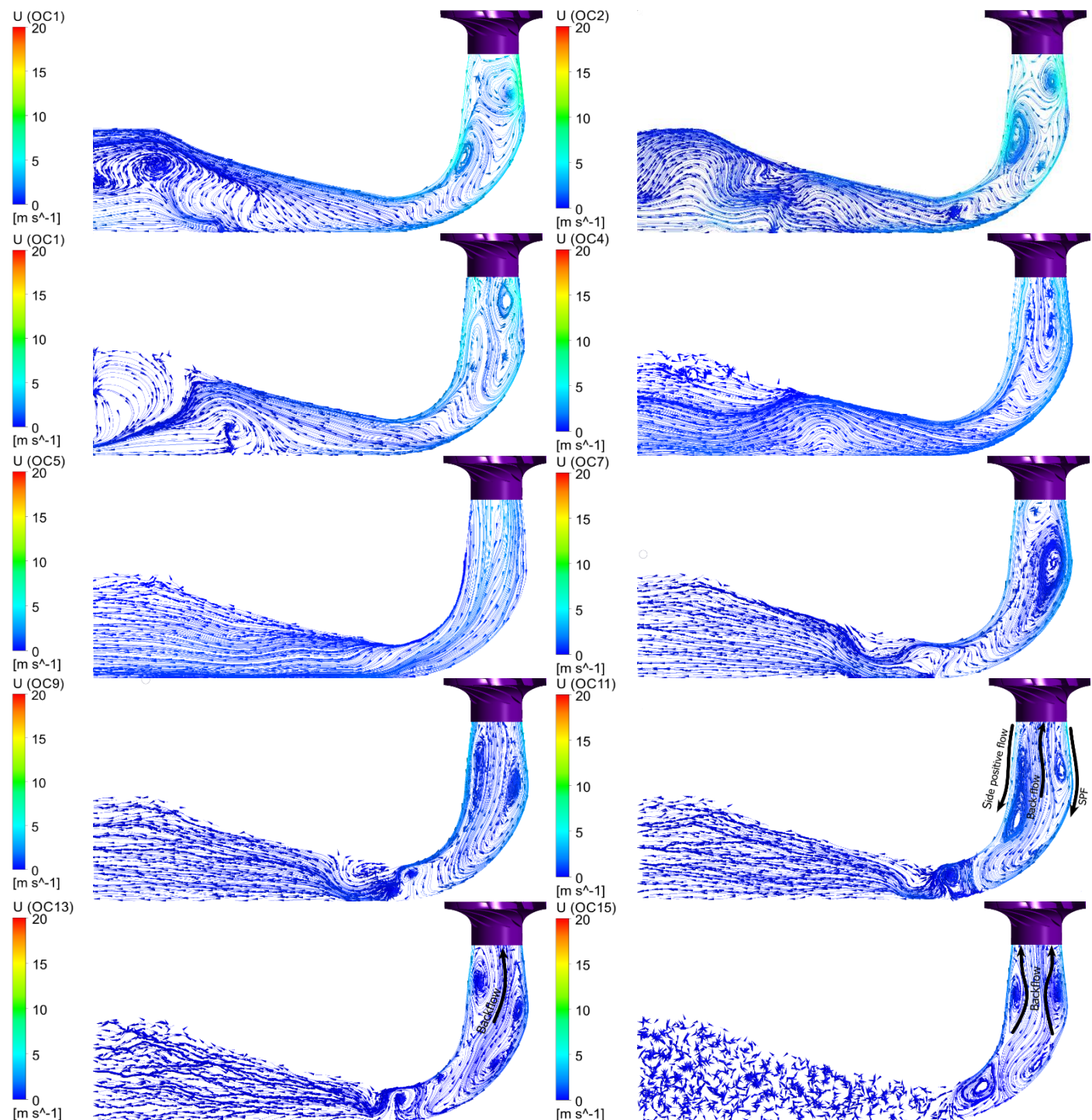

Fig.10 Flow streamlines within the draft tube for different operating conditions

It's worth noting that the mentioned VS flow jet and inter-blade flow vortex are symmetrically and circumferentially distributed along the runner diameter. Under OC2 conditions, the flow is slightly increased while the runner rotating speed is more than doubled, leading to the weakening of the above mentioned VS flow jet and subsequent weakening of inter-blade vortices, where the deep inter-blade channel high-speed flow zone within the BPS vicinities is pushed towards downstream. Under OC3 conditions the above discussed VS secondary flows get weaker, while the former deep inter-blade high-speed flow zone detaches from the 1st blade's pressure side, while the inter-blade BSS-attached flow vortex is considerably weakened on a global scale, before completely disappearing under OC4 conditions, however giving rise to the BPS-attached flow separation and the increase of VS flow speed. The VS flow speed increase is believed to take source from the gradual fast increase in runner rotating speed while the discharge changes by tiny bits. This leads to a faster increase in the tangential component of the incident flow velocity at the runner inlet, which results in the formation of a VS water ring. This water ring gradually gets stronger as the flow decreases, all along to its peak and symmetrically distributed shape under OC15 conditions. As for the cascade flow instability evolution as the machine flow conditions pass through the last phase of the turbine, through runaway all the way down to deep turbine brake zone; one would notice a slight growth of BPS-attached vortex under OC5 conditions, followed by the global shift of inter-blade flow vortices towards the blade leading edge vicinities and the worsening of VS flow vortex. Eventually growing larger and asymmetrically distributed inter-blade flow vortices with subsequent large VS back flows are encountered for the operating conditions in both the runaway vicinities and turbine brake.

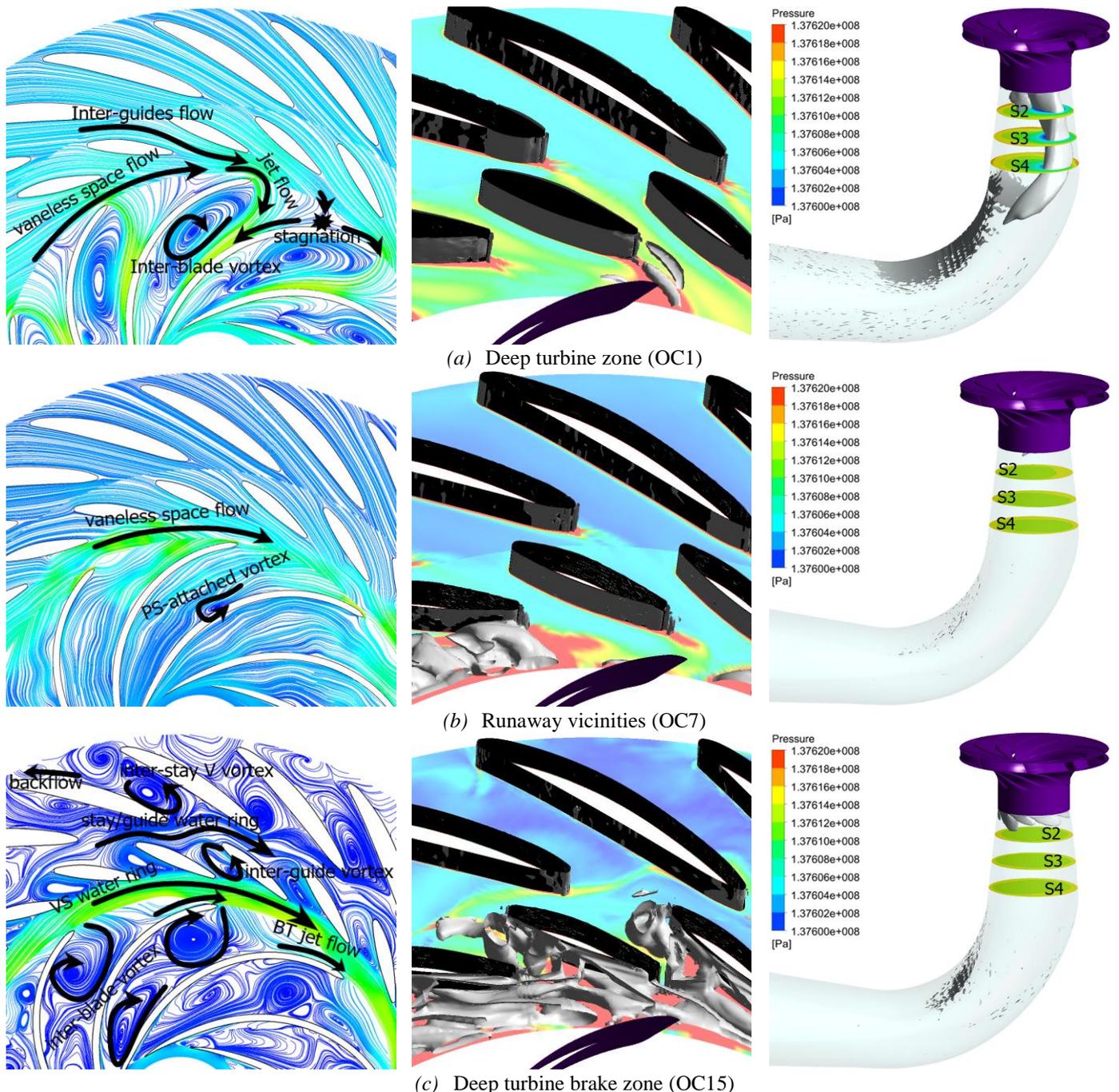

(a) Deep turbine zone (OC1)

(b) Runaway vicinities (OC7)

(c) Deep turbine brake zone (OC15)

Fig. 11 A comparative presentation of flow field characteristics namely, (left) cascade velocity-colored flow streamlines, (middle) cascade flow vorticity, and (right) draft tube flow vorticity.

However, the VS flow instability grows stronger from OC7, getting to its peak under OC9, before weakening towards

OC15. The VS instability is also found to stretch upstream within the guide and stay vane flow channels, as a result of back flow vortices from the vaneless space towards the upstream zones, especially under OC11 and OC13. Under OC15, another vorticity ring is also formed between the stay and guide vanes, leading to the emergence of large vortices within the stay vane channels and subsequent flow reversal back to the scroll casing. It is therefore obvious that as the machine operated under OC1 conditions, larger flow vortices were only situated within the runner inter-blade channels and were symmetrically distributed around the runner diameter. As the machine flow discharge reduced from OC1 to OC4, the inter-blade flow vorticity gradually vanished, marking the zone A as signaled in Fig. 7. Under C5 slight flow separation was notice at every blade's pressure side and vaneless space flow speed started increasing. In the followed operating conditions, flow vertical structures re-emerged within inter-blade channels and got worse as the flow decreased leading to large backflows towards the vaneless and the intensification of vaneless flow instability within the runaway vicinities. In turbine brake, in addition to extreme inter-blade flow vortices that almost blocked the flow passage at a bigger number of channels, flow instability stretched towards the upstream, eventually reaching the stay vane inter-spaces where backflow towards the scroll casing was noticed in a number of channels.

Looking back to flow dynamics and evolution within the runner downstream flow zone, the draft tube flow zone, Fig. 10 shows that the draft flow unsteadiness has gone through remarkable changes as the machine operating conditions gradually changed. Under OC1, which is the most disturbed state of zone A (Fig.7), three vortices can be seen, among which, one is situated within the cone, another in the elbow, and the third within the draft tube outlet vicinities. Throughout the turbine operating zone, draft tube flow unsteadiness continuously decreased with major vortices eventually weakening and disappearing over time, leading to OC5 being the least disturbed of all. However, flow vertical structures re-emerged and grew larger and stronger from the last phase of the turbine operating zone all down to deep turbine brake zone, where draft tube reverse flow grew bigger and bigger reaching its most severe level under OC15. It's therefore obvious that flow instability that took place under certain operating conditions has spread through all components of the investigated computational domain, but with different strengths depending on the component and operating condition under consideration. This is better reflected by a comparative presentation of flow structures in different flow zones under three randomly picked operating conditions as shown in Fig. 11. In this figure, the left portion presents the velocity-colored flow streamlines within the stay/guide vanes and runner inter-blade channels, flow vorticity within the stay/guide vanes in the middle, and draft vorticity on the right side. Free operating conditions, namely OC1, OC7, and OC15 were selected to reflect the occurred big changes in flow structure formations. Under OC1, as also explained in the above sections, flow vortical structures are mainly located within the runner inter-blades channels are symmetrically distributed along the runner diameter. Moreover, comparatively slight flow separations are noticed within the stay/guide vanes due to the endured flow wakes behind the vanes. The latter get even intense within the vaneless space, where together with runner inter-blade backflow, the flow jet structure is formed at every blade's leading edge. Under the same conditions, a vortex rope is noticed within the draft tube from the occurred serious swirling flow within the draft tube flow zone. This is discussed in more details within the next sections. Under OC7 however, the runner inter-blade vortical flow considerably weakens and shift to runner blade's pressure side, while the VS flow gets more disturbed with the formation and eventual strengthening of VS water ring. As for the stay/guide vane zone, the flow unsteadiness almost disappears as the formerly noticed flow wakes vanish. Correspondingly, the draft tube vortex rope weakens to the point of disappearance. Finally, under OC15 conditions, extreme and asymmetrical flow unsteadiness develops within the runner inter-blade channels almost blocking every channel. This leads to huge backflow to the vaneless space where the flow instability not only covers the whole vaneless zone but also expands towards both the guide and stay vane inter-spaces. Under these conditions however, the draft tube does not experience a vortex rope occurrence but flow vortical structures are formed in the wall vicinities next to the runner outlet zone.

3.2. Pressure field characteristics and analysis

In order to get a deep understanding on reversible pump turbine flow dynamics evolution and create a logical link to the correspondingly occurring pressure pulsations at different flow zones, in addition to the above presented discussion on flow field characteristics, a deep and extensive investigation is also conducted on local pressure pulsation characteristics at different flow zones for different operating conditions. This is also in line with this study's objective to investigate the instability transmissibility among different components of the investigated computational domain. To do so, different pressure monitoring points have been positioned at different flow zones along the full machine water flow passage. Fig. 12 shows the positions, number, and names of the utilized pressure monitoring points. 12 monitoring points, namely S1 to S12, were positioned within the scroll casing in the vicinities of the stay vanes' inlet zone, where a constant angular interval of 30° was kept between every two successive monitors (See Fig. 12 (a)). Moreover, as shown in Fig. 12 (b), four equidistantly spaced monitors (ST1 to ST4) were positioned within the inter-stay vanes channels, while 12 more locations were selected to monitor the pressure pulsation characteristics at the guide vane ring inlet zone (G1 to G12). Other 12 monitors were positioned within the vaneless space around the runner inlet zone with 30° spacing between every two successive monitors (R1 to R12). As for the runner inter-blade flow zone (See Fig. 12 (c)), only one inter-blade channel was investigated, where three groups of monitors were selected, and were located at 0.2, 0.6, and 0.9 positions considering the streamwise unitary distance from runner outlet (position 0) to runner inlet (position 1). For each of the three groups (IB1, IB2, and IB3), monitors are positioned in an order that starts from the first blade's suction side towards the second blade's pressure side as shown in Fig. 12 (c) . The first group consists of three monitors namely IB11, IB12, and IB13. Then five monitors have been selected for the second and third groups as IB21-IB25 and IB31-IB35 respectively. Finally, 39 pressure monitoring points have been set within the draft tube on three parallel planes, namely S1, S2, and S3 (See Fig. 12 (d)). On plane S1 which is located at the runner outlet, 12 monitoring points (S11-S112) have been circumferentially set in the draft tube wall vicinities with a repetitive interspace of 30°. Following the same arrangement as on S1, 12 monitors have been considered on each of planes S2 and S3. Moreover, one more point has been set at the center of each of the tree planes, leading to three more central monitoring points (DT1 on S1, DT2 on S2, and DT3 on S3).

To get a clear picture of pressure pulsation characteristics for different operating conditions and possible pulsation frequency components at different machine flow zones, OC1 operating condition was chosen, where using the Fast Fourier Transform method (SFT), pressure pulsation spectra within the machine full flow passage have been deducted. Illustrations in Fig.13 to Fig.15 show the FFT-based pressure pulsation spectra within the vaneless space, its upstream and downstream flow zones.

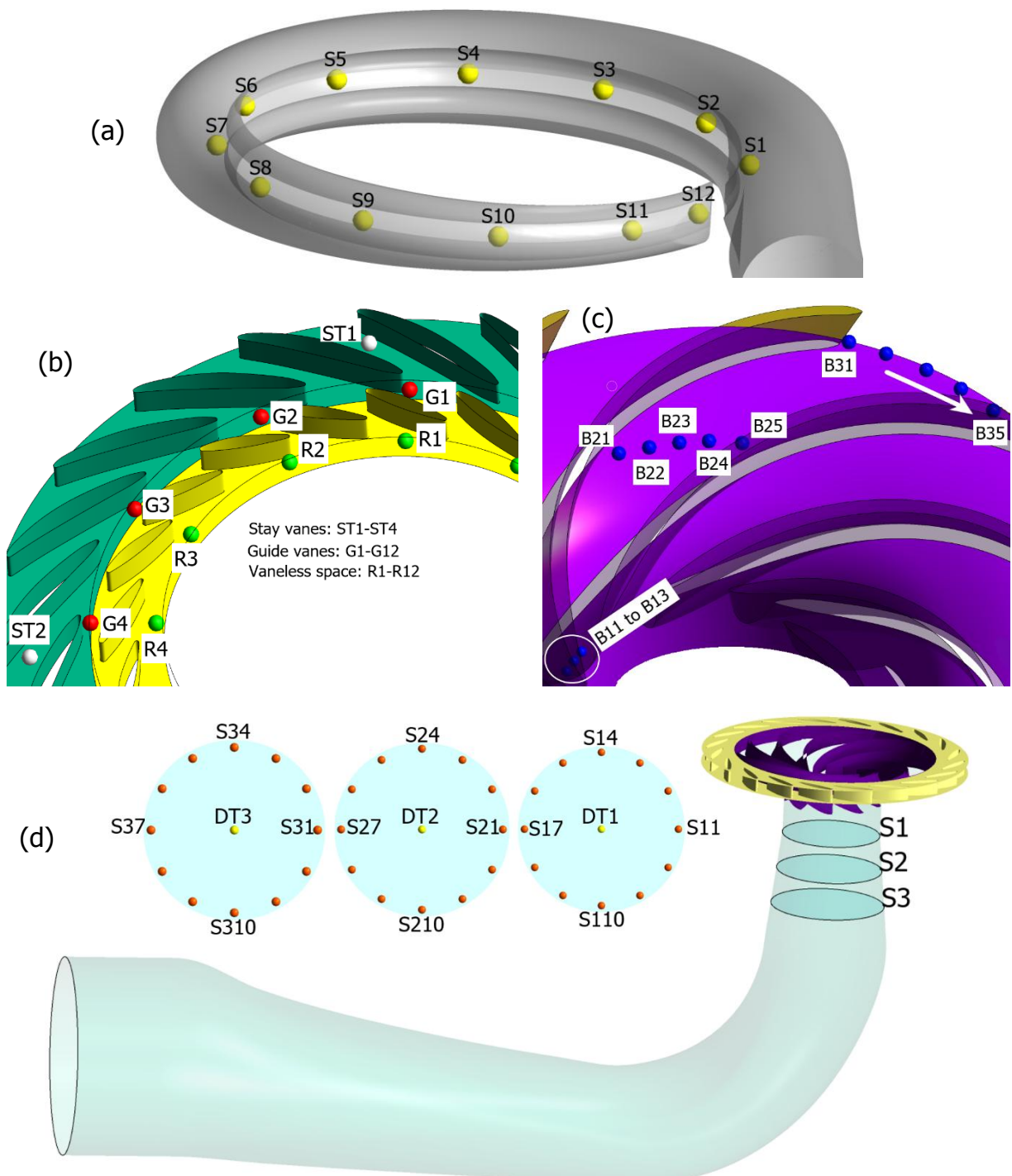

Fig. 12 Position of pressure monitoring points at different flow zones (a) draft tube, (b) Stay/Guide vanes, (c) Runner, and (d) Draft tube

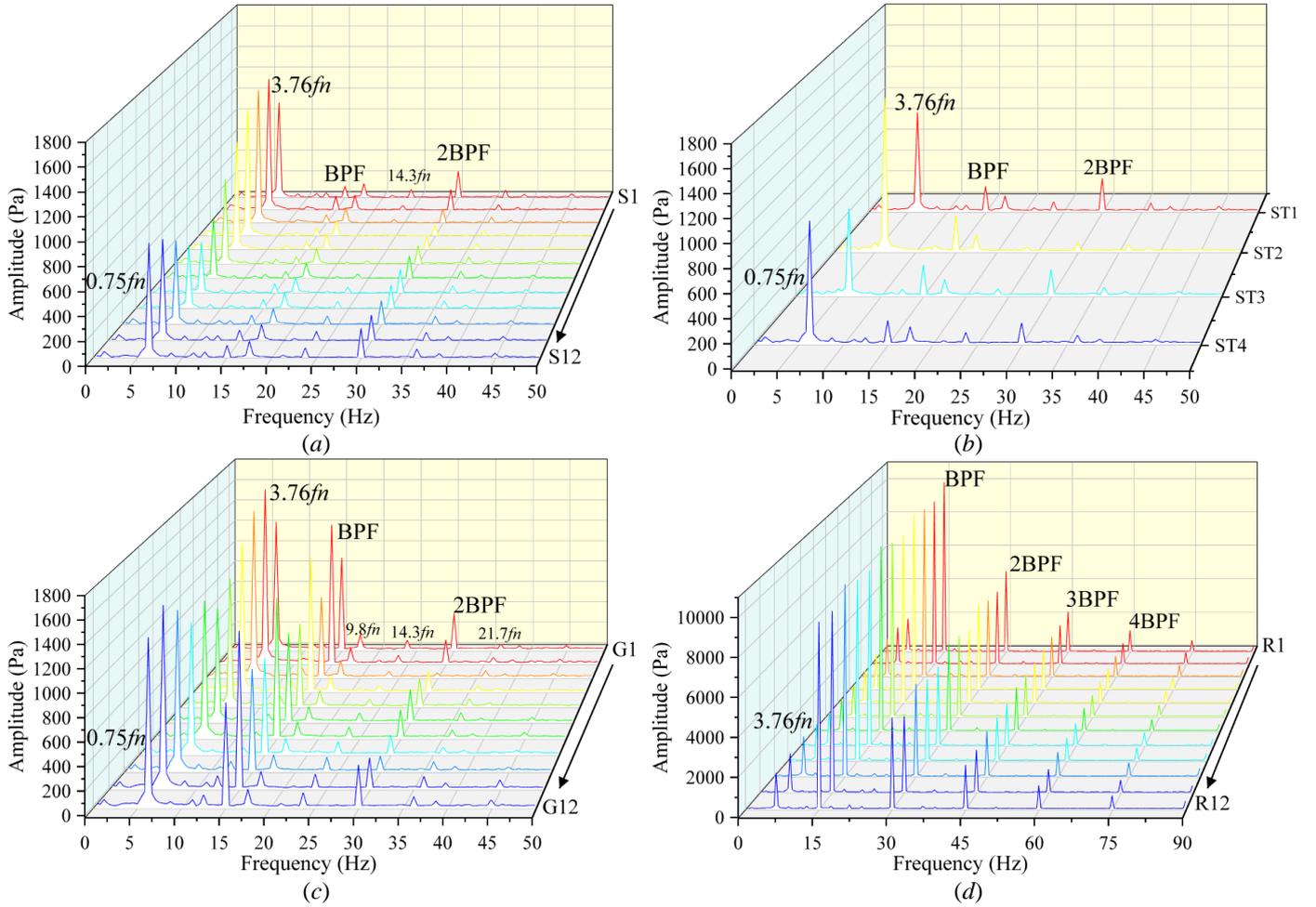

Fig.13 Pressure pulsation spectra for the OC1 operating conditions (a) Scroll casing (b) Stay vanes (c) Guide vanes and (d) vaneless space.

Starting with the upstream flow zone, Fig.13 shows the pressure pulsation characteristics within the vaneless space, guide vane inter-spaces, stay vane inter-spaces and scroll casing. The vaneless space pressure pulsation spectrum (Fig.13 (d)) is found to be constituted by two main components, namely the blade passing frequency (BPF) and harmonics (2BPF, 3BPF …) as well as the low-frequency component (LFC) which is 3.76$f_n$ for this case ($f_n$: runner rotational frequency). Note that the BPF is a result of the rotor-stator interactions (RSI) that take place between the runner blades' leading edges and the guide vanes' trailing edges as explained in the article introductory section, and its expression goes like in Equation 7.

$$\text{BPF} = \frac{(n \times Z_R)}{60} \qquad (7)$$

Where $n$ stands for the runner rotational speed in rpm. In this case, the BPF constitutes the dominant frequency with amplitudes far higher than both its harmonics and the LFC. For instance, its amplitudes are estimated 2 times, 4 times, and 7 times higher than 2BPF, 3BPF, and LFC respectively. LFCs are believed to take source from vaneless incurred flow unsteadiness, which may be original from the same zone or transferred from flow zones in the vicinities. Going towards the VS upstream direction, Fig. 13(c) shows the pressure pulsation spectra for the 12 motoring points within the guide vanes inter-spaces towards their leading edges (guide vane ring inlet zone). Compared to the VS case, pressure pulsation amplitudes tremendously drop to almost 12%. In this zone, in the addition to BPF, 2BPF, and the 3.76$f_n$ LFC, a number of other high frequency-low amplitude components (HF-LAC) have emerged such as 9.8$f_n$, 14.3$f_n$, and 21.7$f_n$. These three are also found within the stay vanes and scroll casing (See Fig.13 (b) and Fig.13 (a)), where their amplitudes are however found to continuously decrease as we advance farther in the upstream direction. With remarkably decreased amplitudes of the RSI-born components (BPF and 2BPF) within the GV flow zone, the 3.76$f_n$ LFC now dominates the whole zone with slightly higher amplitudes than the BPF, but considerably higher than the rest of components. It's also important to note that in this flow zone, a new 0.75$f_n$ LFC has also emerged with the lowest pulsation amplitudes. The latter, in addition to three already mentioned HF-LACs, stretch the whole way to the scroll casing. In the same respect, the 3.76$f_n$ LFC also propagates all the way to the scroll casing where it continues to dominate all other components. In other words, pressure pulsation spectra within the VS and upstream flow zones are constituted by the RSI-born components (BPF and harmonics) and local flow unsteadiness-born components (LFCs and HF-LACs). While the BPF component dominates the whole VS, its amplitude is found to steeply drop towards the upstream direction, leading to the 3.76$f_n$ LFC dominating the totality of upstream flow zones. In addition to that, unlike the RSI-born components case, slight amplitude changes are noticed on LFU-born components in the upstream direction.

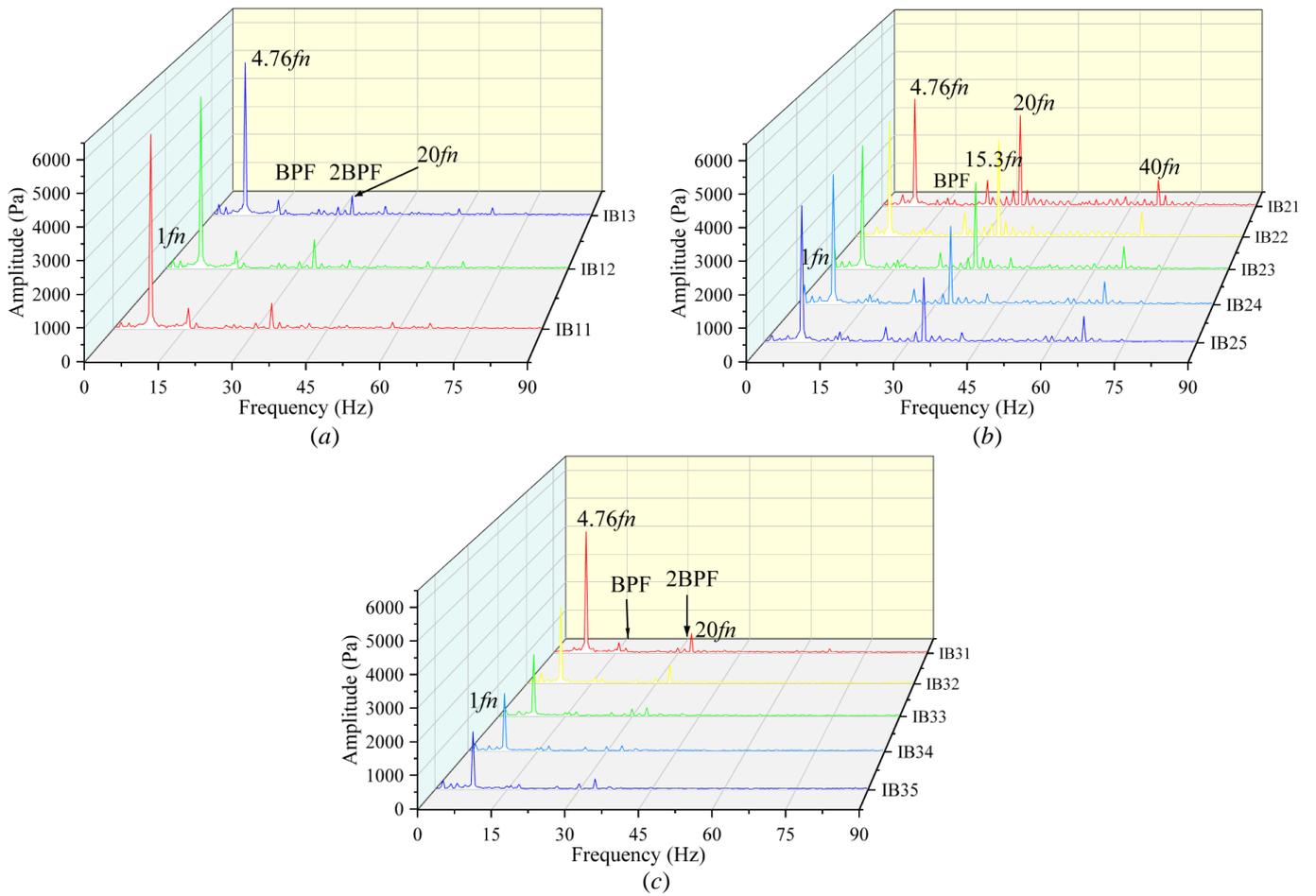

Fig.14 Pressure pulsation spectra for OC1 operating conditions within the runner (a) IB1 (b) IB2 and (c) IB3.

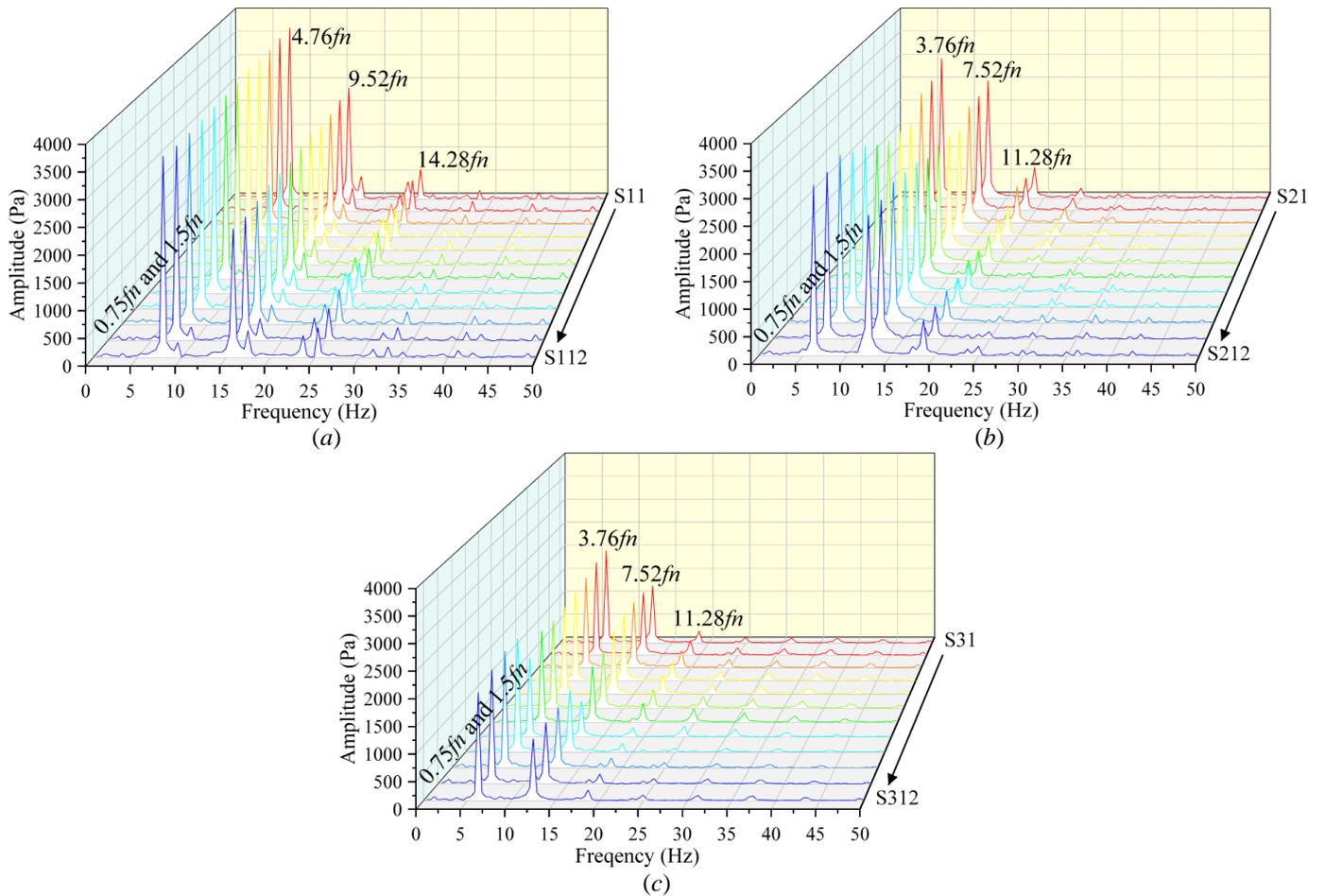

Fig.15 Pressure pulsation spectra for OC1 operating conditions within the draft tube (a) Surface S1 (b) Surface S2 and (c) Surface S3.

As for the VS downstream flow zones, Fig.14 and Fig.15 show the pressure pulsation characteristics within the runner inter-blade flow channels and draft tube respectively. Considering three zones aligned along the meridional direction from runner outlet to inlet and corresponding monitoring point groups as explained in upper sections, it can be seen that there is a global increase of pressure pulsation amplitudes from the inlet zone towards the outlet. Moreover, one would notice that inter-blade pressure pulsation spectra are still made up by the above-mentioned RSI-born and LFU-born components, with the latter being more dominant. In a bit more details, BPF and 2BPF can be found all along the streamwise direction from runner inlet to outlet zones, but with very small (negligible) amplitudes. On the other hand, one of the newly emerged LFCs, 4.76$fn$, dominates the whole inter-blade flow zone followed by a 20$fn$ component of the HF-LAC type. The latter has also showed up in inter-blade channels even for the rest of all tested flow conditions. It's important to note that, in addition to RSI-born BPF and 2BPF components, a number of new LFU-born components have emerged within the inter-blade flow zone. That is to say 15.3$fn$ and 20$fn$ of the HF-LAC type, and 1$fn$ and 4.76$fn$ of the LFC type respectively. It's therefore obvious that among the formerly discussed pulsation components within the VS, only a number of the RSI-born ones managed to propagate to inter-blade flow channels, even then, with extremely low amplitudes (6%). The inter-blade flow zone was on the other hand found to be dominated by some of the newly emerged LFU components such as the 4.76$fn$ component. The latter is also found to dominate the draft tube inlet zone (runner outlet zone), where the 2$^{nd}$ and 3$^{rd}$ dominant frequencies are its 2$^{nd}$ and 3$^{rd}$ harmonics, 9.52$fn$ and 14.28$fn$ respectively (See Fig.15 (a)). Pressure pulsation characteristics within the draft flow zone are shown through Fig.15. Let's keep in mind that, as explained in the above sections, three pressure monitoring planes have been chosen as S1, S2, and S3 in a direction from the draft tube inlet zone towards the elbow; where 12 equidistant monitoring points were circumferentially positioned at each plane's wall vicinal zone. From Fig.15, it can be noticed that, on a global scale, pressure pulsation amplitudes decreased in the direction from the draft tube inlet towards the elbow. Pressure pulsation spectra of monitoring points on Plane S1 partially show the inter-blade pressure pulsation characteristics where the same 4.76$fn$ component still serves the dominant frequency. However, just as it is the case for the rest of draft tube zones, the RSI type pulsation completely disappears throughout the whole draft tube flow zone. Unlike the draft tube inlet situation, the formerly discussed vaneless space LFC (3.76$fn$) re-emerged within the downstream flow zones serving the dominant frequency, where its 2$^{nd}$ and 3$^{rd}$ harmonics were the 2$^{nd}$ and 3$^{rd}$ dominant frequencies (See Fig.15 (b) and Fig.15 (c)). In addition, two new LFCs namely 0.75$fn$ and 1.5$fn$ have emerged on the three investigated planes, where one of them (0.75$fn$) happens to have also appeared within the VS and upstream flow zones. <u>At this stage, draft tube pressure pulsation characteristics, especially on planes away from the runner outlet, are characteristic to a draft tube vortex rope occurrence. The latter is found to rotate at approximately 0.75$fn$ in the same direction as the runner.</u>

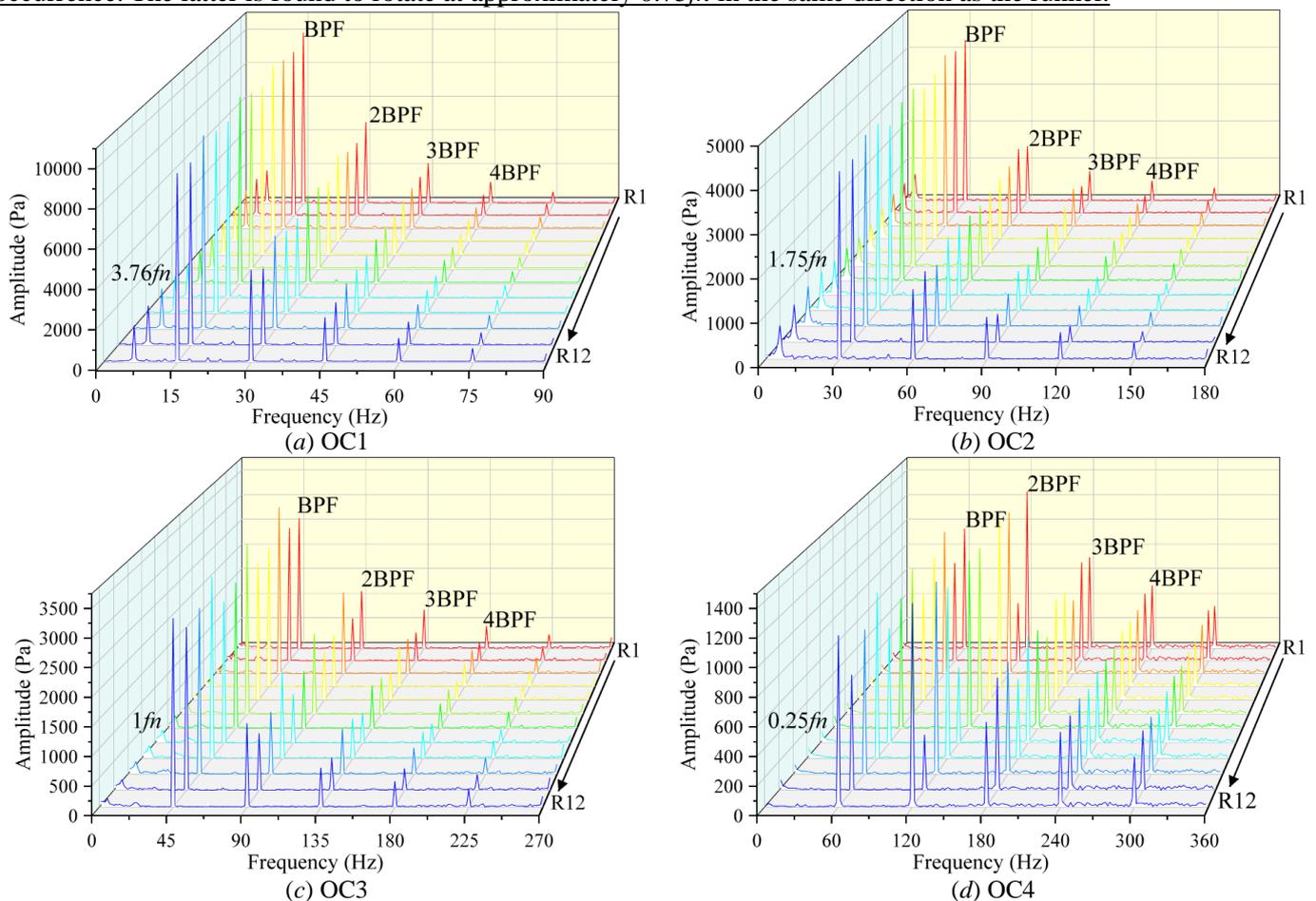

Fig.16 Pressure pulsation spectra within the vaneless space for operating conditions in Zone A.

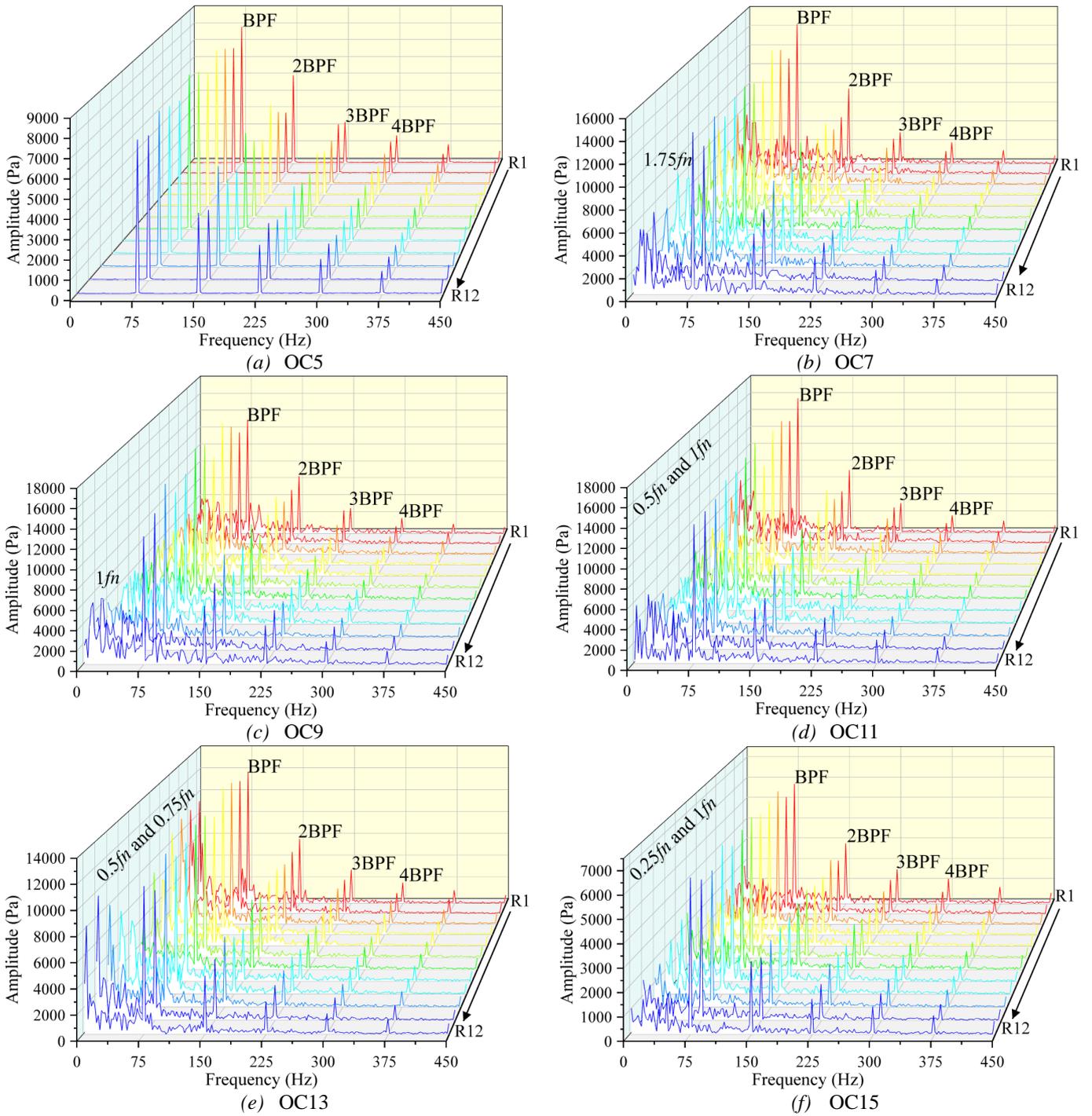

Fig.17 Pressure pulsation spectra within the vaneless space for operating conditions in Zone B.

Having clearly observed that VS flow zone constitutes the base for RPT large pressure pulsations, where individual pulsation frequency components have the ability to propagate to other flow zones both in up and downstream directions; one would predict that VS pressure pulsation characteristics and distribution mode may vary with different operational or machine design-related parameters. That would also reflect corresponding modifications in terms of pressure pulsation propagation mode and characteristics to other flow zones. This aspect constitutes the main objective of the present study, where VS pressure pulsation characteristics and its transmissibility to other flow zones are first studied, before investigating the effect of two parameters, namely the machine influx ($Q$) and runner structural design (blade number specifically), on the same. To do so, VS pressure pulsation characteristics for different machine flow conditions are first investigated. Fig.16 and Fig.17 demonstrate variations of VS pressure pulsation characteristics for ten different flow conditions classified into zones A and B respectively. As also discussed in the above sections, Zone A represents all operating conditions where large symmetrically distributed flow vortex is found within the runner inter-blade flow channels. It has been demonstrated that the latter weakens with the gradual decrease in machine influx $Q$ and corresponding increase in rotational speed $n$. In this zone as shown in Fig.16, the VS pressure pulsation amplitudes are globally found to substantially drop as the machine flow deceases, where the RSI-born BPF component is the dominant frequency for the majority of flow conditions, at the exception of OC4 conditions where BPF's 2$^{nd}$ harmonic (2BPF) serves the dominant frequency.

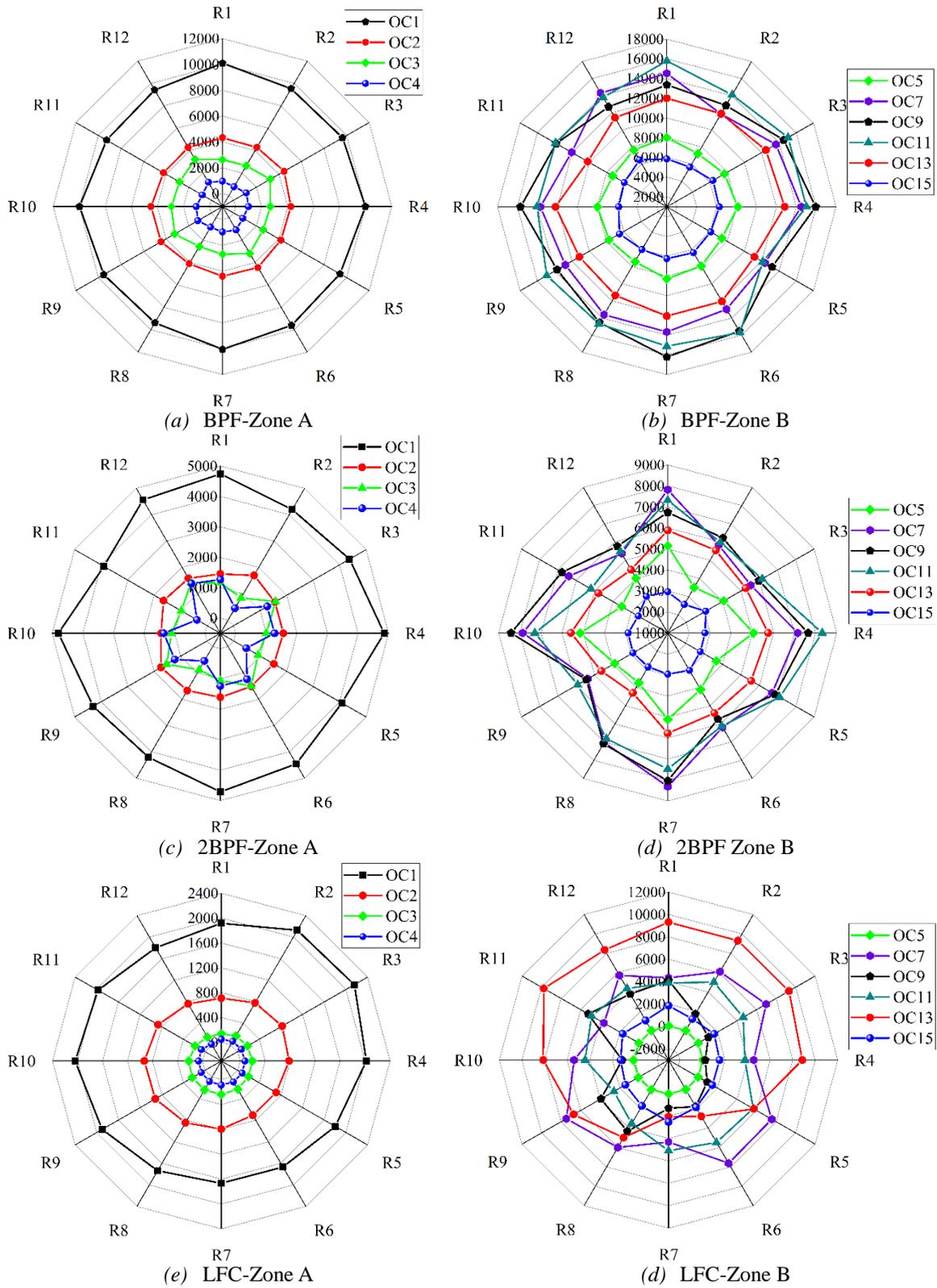

*(a)* BPF-Zone A
*(b)* BPF-Zone B
*(c)* 2BPF-Zone A
*(d)* 2BPF Zone B
*(e)* LFC-Zone A
*(d)* LFC-Zone B

Fig.18 Vaneless space pressure pulsation distribution mode

In addition, in line with the experienced gradual increase in rotational speed, the value of BPF has constantly increased for each flow conditions, leading to 15Hz, 30Hz, 45Hz, and 60Hz being BPF values for the four flow conditions in Zone A respectively. On the other hand, in line with the continuous weakening of flow unsteadiness as the machine flow conditions changed from OC1 all the way to OC4, the value of LFCs has gone through a gradual decrease from 3.76$f_n$, through 1.75$f_n$ and 1$f_n$, to 0.25$f_n$ respectively. From OC4, the VS pressure pulsation amplitude has globally increased towards the runaway vicinities before dropping towards deep turbine brake zone operating conditions.

Throughout Zone B, the BPF continues to be the dominant frequency, where the 2$^{nd}$ dominant frequency could either be BPF's 2$^{nd}$ harmonic or one of the LFCs depending on the concerned operating condition. Moreover, due to slight changes in runner rotational speeds in zone B, the BPF value has been almost constant throughout the whole zone. Though LFCs were absent (comparatively negligible amplitudes) for OC5 conditions, their numbers and amplitudes increased with extreme flow instability that took place within the VS, upstream, and downstream flow zones as machine flow conditions passed through zone B. unlike zone A situation, LFCs in zone B were many, with slightly varying amplitudes, which made it hard to dissect the dominant ones for certain conditions. However, a number of LCs such as 0.25*fn*, 0.5*fn*, 0.75*fn*, 1*fn*, 1.75*fn* have been observed at different operating conditions.

Fig. 18 shows the VS pressure pulsation distribution mode through its most remarkable components, namely BPF, 2BPF, and LFC. On a global scale, it can be seen that pressure pulsation in Zone B is always far larger than Zone A regardless of the considered frequency component. Due to a comparatively low level of flow unsteadiness within Zone A, pressure pulsations within the VS are found to almost distribute symmetrically with respect to the rotational axis, which is not the case for Zone B, especially with the LFCs distribution mode.

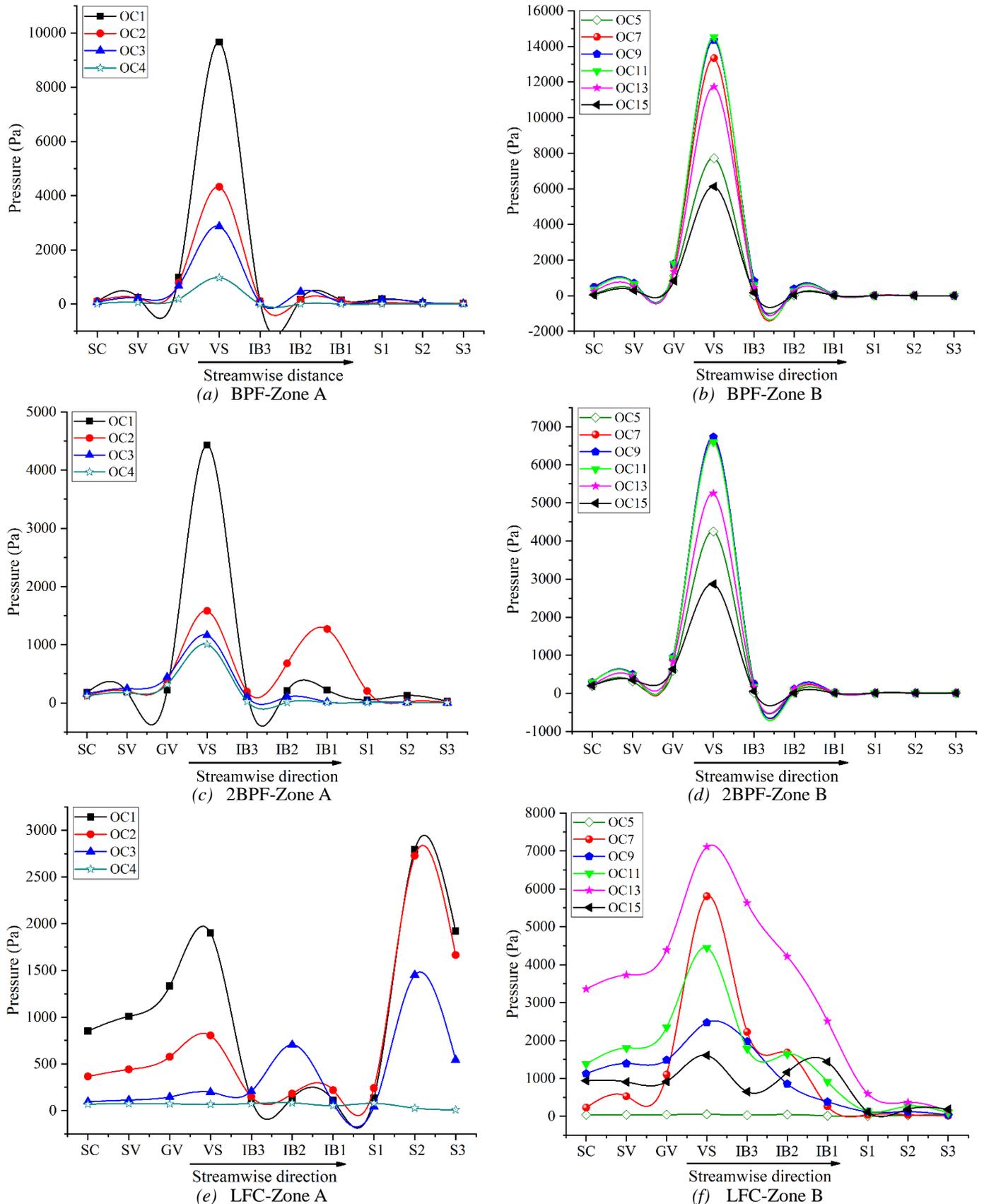

Fig. 19 Transmissibility of main pulsation frequencies among different flow zones

While pressure pulsation amplitudes are found to decrease with the machine flow conditions in Zone A, the situation in Zone B presents a different trend, where pulsation amplitudes are generally found to increase from OC5 to runaway vicinities, before falling back to deep turbine brake zone. Note that, though they exhibited the lowest level of pulsations, thus presenting a negligible effect on VS global pulsation distribution mode, the LFCs did not follow the above stated trend where the highest of pressure pulsations are found with OC13 while the lowest is with OC5 rather than OC15. This takes source from the fact that LFCs are LFU type, thus relate very much with instantaneous flow instability at specific flow zones. In this respect, as also demonstrated in Fig.9, OC5 presents the most stable flow structure in Zone B, thus presenting the lowest level of LFC amplitudes. Overall, throughout the whole range of ten investigated flow conditions, BPF pulsation amplitudes are the highest thus being a controlling factor, where global VS pressure pulsation variation profile almost correlate with BPF evolution profile. In this figure,

In order to investigate the propagation characteristics of the main pulsation frequency components (BPF, 2BPF, and LFC) as discussed in the above sections, correspondingly recorded pulsation amplitudes within each of the five components of the investigated computational domain have been averaged to come up a streamwise distribution mode for each of the components under different operating conditions. Fig. 19 is therefore a presentation of the pulsation amplitudes variation as the flow passes through the RPT full flow passage from the scroll casing (SC) through Stay/guide vanes (SV and GV) down to the vaneless space (VS) and inter-blade channels, all the way down to draft tube flow zone. In this figure, some of truths presented in the above sections are reconfirmed where for instance, regardless of the considered frequency component, the VS flow zone exhibits the highest of pulsation amplitudes. Both BPF and 2BPF components, globally being the $1^{st}$ and $2^{nd}$ dominant frequencies, their amplitudes have been found to decrease from OC1 to OC4 before increasing to OC9 and OC11(runaway vicinities) to finally drop again towards OC15. This is displayed by both Fig. 18 and Fig.19. Regarding their ability to be transmitted to up or downstream flow zones, Fig. 19 shows that both BPF and 2BPF can be easily transmitted to upstream flow zones with critically damped amplitudes reaching the scroll casing flow zone. Nevertheless, in the VS upstream zones, the operating conditions in Zone B are found to exhibit higher amplitudes than Zone A, where the above presented operating condition-based pulsation profile is respected. That's to say that pulsation amplitudes decreased with machine flow conditions for zone A, while they first increased then decreased for zone B. For the downstream flow zones, the transmissibility of the $1^{st}$ and $2^{nd}$ dominant frequencies varies depending on the considered flow condition. For instance, starting from the VS zone, pulsation amplitudes in Zone A are globally found to steeply drop to almost 5% at the runner inlet zone (IB3), and continuously decrease downstream to runner outlet zones (S1), from where they become null all through downstream flow zones. For zone B however, the stated steep fall from VS to IB3 zone is still noticed, followed by a complete vanishment of both components within the inter-blade zone and draft tube for all flow conditions (from IB1 towards downstream).

It's also important to note that, unlike other flow conditions, the amplitudes of both BPF and 2BPF became null right at the runner inlet (IB3) for OC4, OC5, and OC15 conditions respectively. On the other side, LFCs are found all the way to VS up and downstream flow zones, where for flow conditions in Zone A, the lowest of pulsation amplitudes are within the inter-blade channels while the draft tube zone recorded the highest level of pulsation amplitudes, even higher than the VS zone. This is linked to the formally mentioned draft tube vortex rope that took place under OC1conditions and weakened until its disappearance under OC3 conditions. Besides, LFCs are found to expand also to VS upstream, where unlike the RSI-born frequency components, the amplitudes are far less damped. For the whole zone A, OC1 presented the highest pulsations especially in VS upstream and draft tube zones, while OC4 exhibited the lowest level of pulsations throughout all machine components. On the other hand, Zone B globally presented higher pulsations than Zone A. In this group, unlike the Zone A situation, lowest pressure pulsations were noticed within the draft tube. From the VS zone, pulsation amplitudes gradually decreased towards up and downstream directions, where unlike the RSI-born components, the highest and lowest amplitudes were recorded with OC13 and OC5 flow conditions respectively. Overall, VS pressure pulsation is the highest of all flow zones, where its two main RSI-type pulsation frequencies (BPF and 2BPF) are found to propagate to upstream flow zones reaching the scroll casing with critically damped amplitudes. However, in the downstream direction, these frequencies are quickly damped were the farthest they could travel before vanishing was the runner outlet flow zone. On the other hand, LFCs are found to propagate to both up and downstream with slightly damped amplitudes. The noticed draft tube high pulsation amplitudes for the first three flow conditions within Zone A, are linked to the occurred draft tube rotating vortex rope.

3.3. Influence of runner blade number

In order to investigate the effect of runner design on the above discussed truths, the runner blade number has been gradually decreased from 9 through 8 to 7 blades, where the machine flow dynamics for each of the runner models were simulated and analyzed, considering two selected flow conditions namely OC1 and OC11. These two conditions, as also presented in the above sections, represent the highest flow condition in turbine mode and one of two investigated conditions within the runaway vicinities respectively. Under these two conditions, as also pointed out in the above sections, the machine endures the highest of pressure pulsations, where OC1 is selected from Zone A and OC11 from Zone B. Keeping in mind that this investigation focuses on RPT's VS pressure pulsation characteristics and their transmissibility to both up and downstream flow zones, the next figure, Fig. 20, displays the VS pressure pulsation amplitudes distribution mode and the effect of runner blade number on the same.

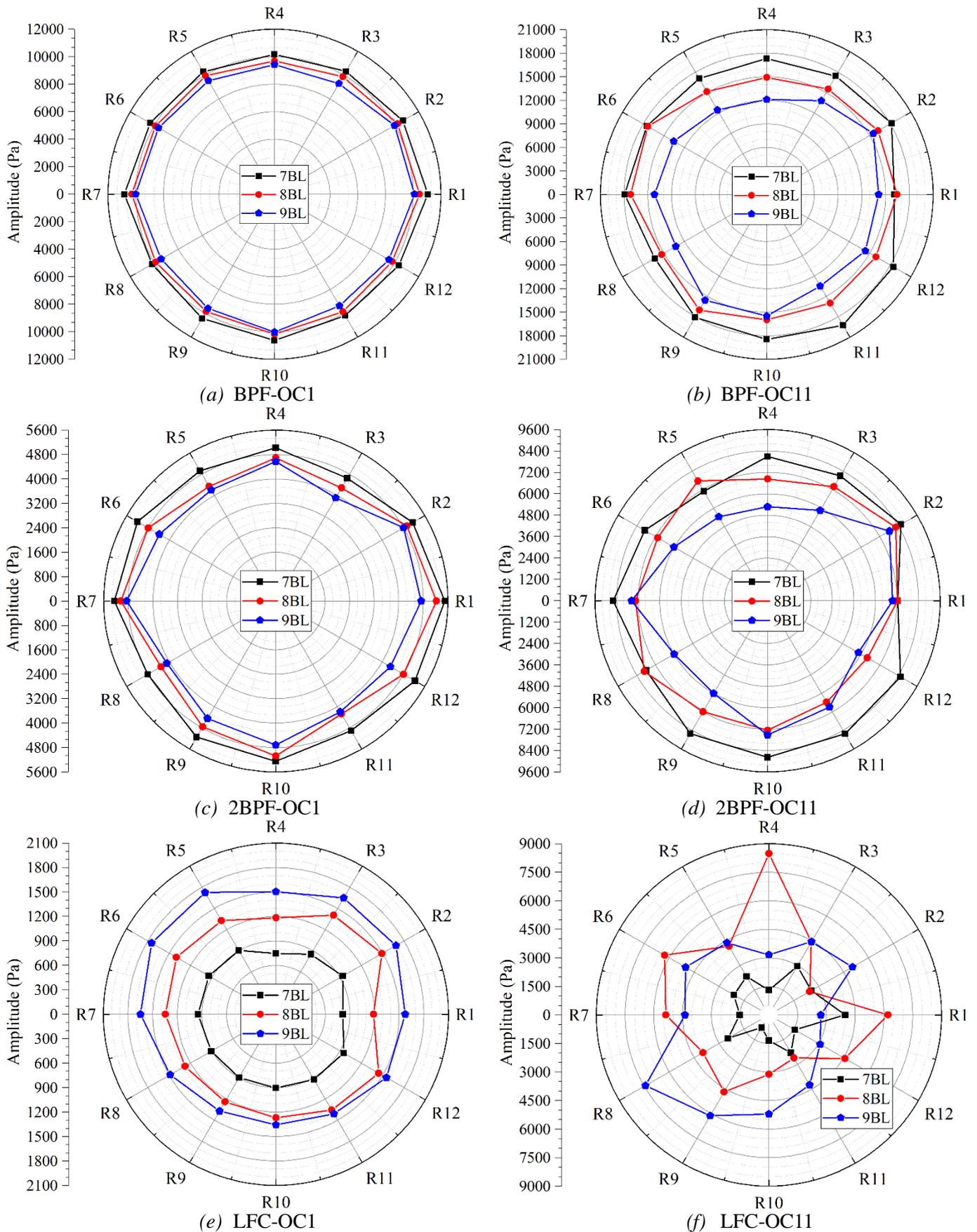

Fig. 20 VS pressure pulsation distribution mode for different runner blade numbers

In this figure, it is shown that, regardless of the considered blade number, the RSI-born BPF components still dominates the whole vaneless space, followed by its 2nd harmonic for which pulsation amplitudes are almost half the BPF's. Moreover, pulsation amplitudes distribution mode is found to be symmetric with respect to the runner rotational axis for OC1 conditions, especially the BPF component, while it get distorted under OC11 conditions with the LFC component presenting the worst case. This has something to do with considered operating condition and associated flow dynamics. For instance, while the OC11 is marked by serious and asymmetric inter-blade vortices that instantly block a number of flow channels leading to huge backflows towards the vaneless space, OC1 enjoys a quite small load of symmetrically distributed inter-blade vortices, that are mostly attached on the blade suction side, causing the flow obstruction on the next blade's pressure side, but with little to no backflow to vaneless space. Therefore, VS pressure pulsation characteristics under OC1 are only a result of RSI and a symmetrically distributed LFU-born pressure pulsation. Note that this LFU is linked with the occurred flow wakes that took place at every blade's leading edge, owing to large flow velocity and resultant high incidence angle. Extreme VS asymmetric pulsation distribution as the one for LFCs under OC11 showcase a highly disturbed flow regime where, as shown in Fig. 20 (e) and (f), pulsation amplitudes for LFCs under OC11 are 3 to 4 times higher than OC1. In general, considering the effect of blade number vaneless space pressure pulsation characteristics, Fig. 20 has globally shown that, regardless of the considered RSI-born pulsation component, where BPF would otherwise be given priority, pressure pulsation amplitudes within the vaneless space zone has been gradually decreasing with the increase in runner blades number for both the investigated flow conditions. On the other hand, with the increase in runner blades number, pressure pulsation amplitudes of the LFU-born LFCs have correspondingly increased for both the investigated flow conditions. Though LFC amplitude distribution mode under OC11 conditions is a bit complex especially with the cross-cutting of 8BL and 9BL distributions, but one would notice that 9BL surpassed 8BL amplitude at six locations (R8-R11, R2, and R5), while the opposite only happened at 5 locations (R1, R4, R6, R7, and R12) leading to holding the above stated hypothesis. This may be linked to the fact that the increase in runner blade number makes the runner inter-blade channels smaller in size, which would increase the flow obstruction at these zones, leading to more backflows towards the vaneless space and a subsequent increase in VS pressure pulsation levels. In Fig. 21, an effort has been made to observe the effect of the above VS pressure pulsation changes to both the up and downstream flow zones. In the downstream direction, considering OC1 flow condition, BPF pulsation amplitudes are found to nullify at the runner inlet for 7BL and 9BL, while 8BL dominates the runner inter-blade flow zone, followed by draft tube negligible amplitudes for the three models. The same 8BL dominates the upstream zone with 9BL presenting lowest level of pulsation in the same zone. As for OC11 condition, the VS amplitudes trend holds even at upstream zones, while the downstream is dominated by 7BL especially within the runner, followed by a complete nullification at the runner exit zone (IB1).

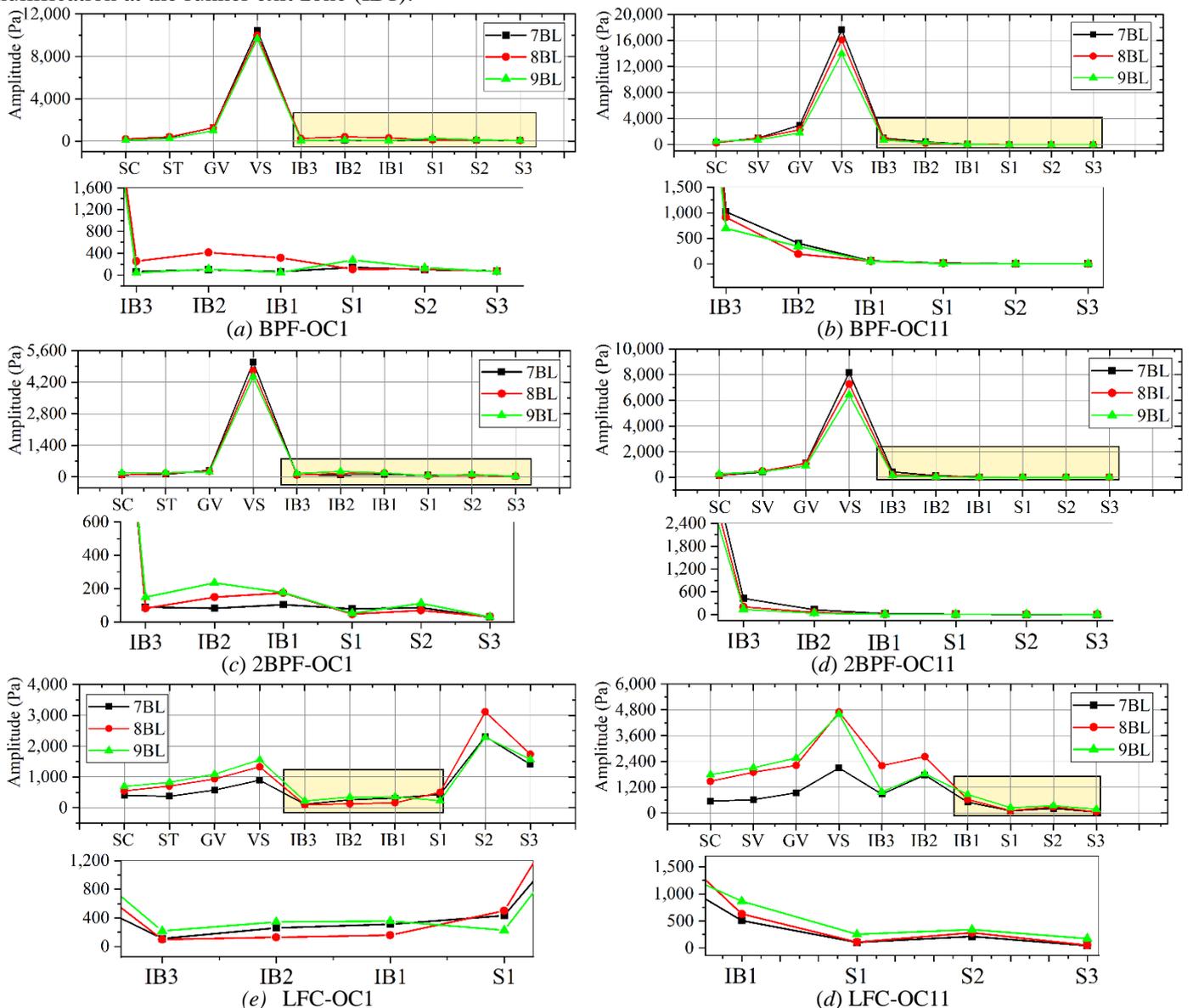

Fig. 21 Transmissibility of main pulsation frequency components for different blade numbers.

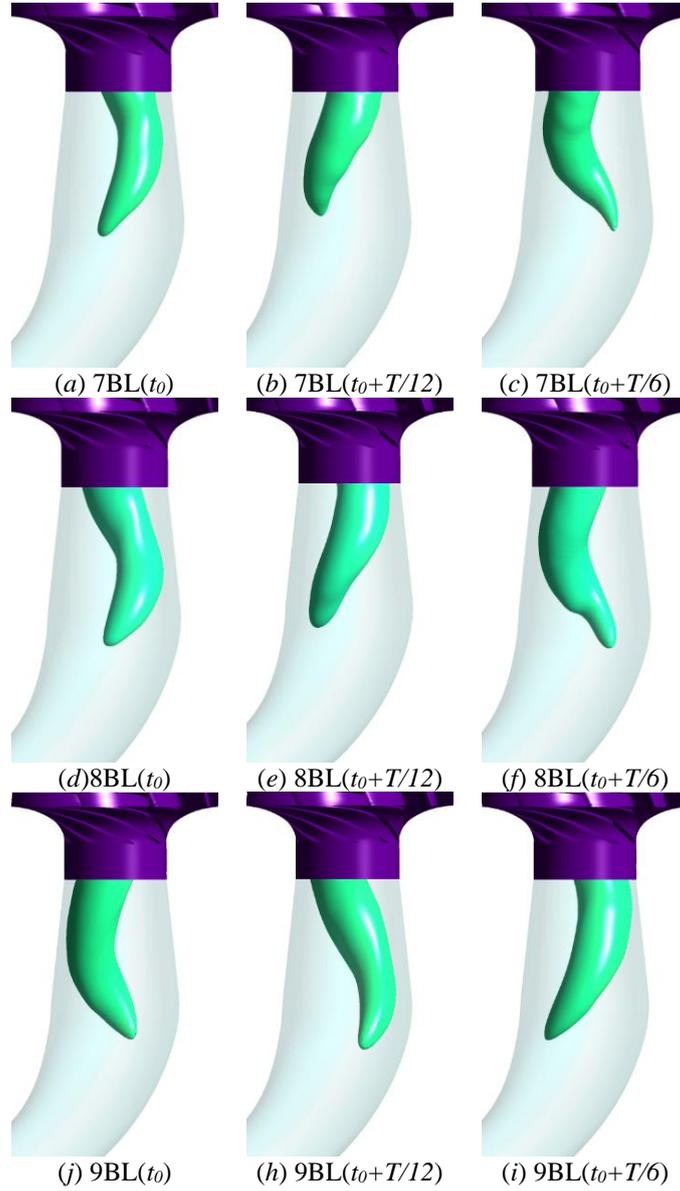

(*a*) 7BL(*t₀*)    (*b*) 7BL(*t₀+T/12*)    (*c*) 7BL(*t₀+T/6*)

(*d*) 8BL(*t₀*)    (*e*) 8BL(*t₀+T/12*)    (*f*) 8BL(*t₀+T/6*)

(*j*) 9BL(*t₀*)    (*h*) 9BL(*t₀+T/12*)    (*i*) 9BL(*t₀+T/6*)

Fig. 22 Evolution of draft tube vortex rope for the three blade numbers under OC1 flow conditions

The 2BPF component, under OC1 conditions, exhibited almost equal pulsation levels for the three models in the VS upstream zone while runner flow zone was dominated by 8BL before nullifying at the draft tube inlet (S1). The same happened with OC11 conditions, where however the runner flow zone was dominated by 7BL, nullifying at the runner exit zone.

As for LFCs, regardless of the runner blades number, the highest pulsation amplitudes are recorded within the draft tube for OC1 conditions, while the same flow zone recorded the lowest level of pulsations under OC11 conditions. While the above discussed VS blades number-based amplitudes trend holds in the whole upstream zone (pulsation amplitude increases with blade number), downstream zone is a bit complex. In this zone, 9BL and 8BL dominate the runner and draft tube zones for OC1 conditions respectively, while the opposite takes place under OC11 conditions. The LFC situation goes in complete correspondence with the locally occurred flow instability in line with the operating condition of concern. For instance, the noticed large pressure pulsations within the draft tube zone under OC1 are more linked to the emerged rotating vortex rope (See Fig. 22), while the VS ones under OC11 conditions are linked to VS large flow instability in form of rotating stall. Keeping in mind that RPT draft tube pressure pulsation is a combined effect of VS-propagated LFCs and draft tube local flow unsteadiness, an attempt has been done to investigate the effect of runner blades number on draft tube vortex rope under OC1 conditions. As shown in Fig. 22, and in partial agreement with Fig. 21 (e), one would roughly notice that draft tube vortex rope weakens with the decrease in runner blade number.

## 4. Conclusion

In this study, an investigation has been carried out on the transmissibility of vaneless space flow instability to both up and downstream flow zone, considering the effect of both flow (10 flow conditions) and runner blades number (3 blade numbers) for an RPT under large GVO (34mm). CFD-backed numerical simulations have been carried out for the stated conditions, where the following concluding remarks have been finally drawn:

a) RPT Vaneless space pressure pulsation is a result of a combined effect of rotor-stator interaction (RSI) and the machine operating conditions-dependent local flow unsteadiness (LFU). The latter is found to take source from vaneless space flow separations as a result of flow wake formation at each blade's leading edge under large flow conditions, while the rotating stall from large asymmetrically distributed flow vortices within inter-blade, vaneless space, and upstream flow zones constitutes the trigger under low flow conditions, especially within the runaway vicinities.

b) With instant changes in machine flow conditions, VS pressure pulsation amplitudes and distribution mode are found to correspondently change. They continuously decreased as the machine influx decreased (Zone A) before rising towards its peak values under runaway vicinal conditions (OC9 and OC11), followed by a gradual fall to deep turbine brake flow conditions (OC15). The transmissibility of VS's both RSI-born (BPF and 2BPF) and LFU-born (LFCs) pulsation components to other flow zones has been also found to depend on machine flow conditions, where flow conditions in zone B are found to transmit higher pulsation amplitudes towards the upstream flow zones as compared to flow conditions in Zone A. For LFCs in particular, VS pulsation components are found to propagate to up and downstream zone with far less damped amplitudes as compared to RSI-born components, where draft tube flow zone recorded the highest pulsations under Zone A flow conditions, whereas the same zone recorded the lowest level of pulsations under Zone B flow conditions.

c) The increase of runner blade number has also been found to alter the vaneless space pressure pulsation amplitudes and distribution mode. Starting from the RSI-born components, with an increase of runner blade number from 7 through 8 to 9, VS pressure pulsation amplitudes have been found to correspondingly decrease, where their distribution mode is more symmetric under Zone A than Zone B flow conditions. This however differs from LFCs case where VS pulsation amplitudes have increased with the runner blade number. In line with the effect of blades number on VS pressure pulsation propagation mode considering RSI-born components, the VS's blade number-based changes have been kept at most of up and downstream flow zones. As for LFCs, the VS trend has been kept in upstream flow zones, while downstream zones recorded different trends between high and low flow conditions.

From the above presented details, it is obvious that both operational and machines structural design parameters considerably influence the RPT flow dynamics and subsequent pressure pulsation levels and distribution modes. For instance, the presently investigated case showed that high flow conditions gave rise to draft tube vortex rope while low flow conditions gave the VS rotating stall. Both phenomena are widely known to cause instability problems in hydraulic turbines. This study has however investigated the VS-born instability and its propagation to the flow zones, which for instance means that draft tube pressure pulsation characteristics should be a combination of propagated components and local flow unsteadiness-born components. In this study, both the flow runner blade number have found to affect the RPT draft tube vortex intensity. More investigations are still needed to explain the involved effect mechanism.

Acknowledgement

This study was supported by National Natural Science Foundation of China (52009033; 52006053), Natural Science Foundation of Jiangsu Province (BK20200509; BK20200508), and the Fundamental Research Funds for Central Universities (B210202066). The support of Hohai University, China is also gratefully acknowledged.

| *Acronyms* | |
|---|---|
| RPT | Reversible Pump Turbine |
| PSP | Pumped Storage Plant |
| GVO | Guide Vane Opening |
| VS | Vaneless Space |
| NRE | New Renewable Energy |
| BPF | Blade Passing Frequency |
| HF-LAC | High Frequency-Low Amplitude Components |
| BPS | Blade Pressure Side |
| IEC | International Electrotechnical Commission |
| OC | Operating Condition |
| RANS | Reynolds Averaged Navier-Stokes equations |
| SST | Shear Stress Transport turbulence model |
| GGI | General Grid Interface |
| LFC | Low Frequency Component |
| FFT | Fast Fourier Transform |
| LFU | Local Flow Unsteadiness |
| RSI | Rotor-Stator Interactions |

| *Latin & Greek letters* | |
|---|---|
| $Q_{11}$ | Unitary Discharge |
| $T_{11}$ | Unitary torque |
| $n_{11}$ | Unitary rotational speed |
| $y^+$ | Dimensional height from the wall |
| $Q$ | Discharge |
| $H$ | Head |
| $P$ | Static pressure |
| $z$ | Number of runner blades |
| $fn$ | Runner rotational frequency |
| $n$ | Runner rotational speed |
| $g$ | Gravitational acceleration |
| $\eta$ | Efficiency |
| $\rho$ | Density |